\documentclass{article}
\usepackage{graphicx, hyperref, color}
\usepackage{tikz}
\usetikzlibrary{patterns}

\def\bb#1{\textcolor{magenta}{#1}}

\title{Scale Prevention  By Ceramic Balls}
\author{
Hideo Kawarada\footnote{\emph{H. Kawarada has provided the modelling of the problem}  :  74-2, Sanjinsawa, Shimokajiro, Onahama. Iwaki city, Fukushima Japan. 
Postal No. 970-0316.
Fax No :  +81 246 92 2352.
Email: \emph{kawarada0@nifty.com }. }~ and~ Olivier Pironneau \footnote{\emph{olivier.pironneau@gmail.com } has provided the numerical simulations.}
}

\newtheorem{remark}{Remark}

\begin{document}

\maketitle

\tableofcontents
\newpage
\begin{abstract}
In industrial facilities and household equipments scale formation leads to reduced efficiency
and  damages. Therefore various devices for anti-scale have been designed for a long time.  Recently one of them, an aggregation of ceramic spheres, was proposed to prevent scale formation and its efficiency shown experimentally.
The purpose of this paper is to clarify the function of this device by proposing several mathematical models and by pursuing their mathematical and numerical analysis. The first  model measures the effect on nucleation of calcite of the electric potential near the surface of a ceramic sphere in natural water. The second model is based on a crystal lattice model and argues that the surface tension energy of the calcite particles is reduced by the polarization energy brought by the ceramic balls.  The third model is macroscopic and numerical and studies the effect of two ceramic balls arrangements in a flow of water containing calcite particles.   
\end{abstract}



\section{Introduction}
\begin{figure}[htbp]
\begin{center}
$$\includegraphics[width=7.8cm]{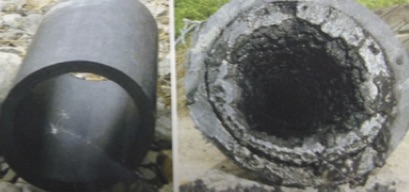}~
\includegraphics[width=4cm]{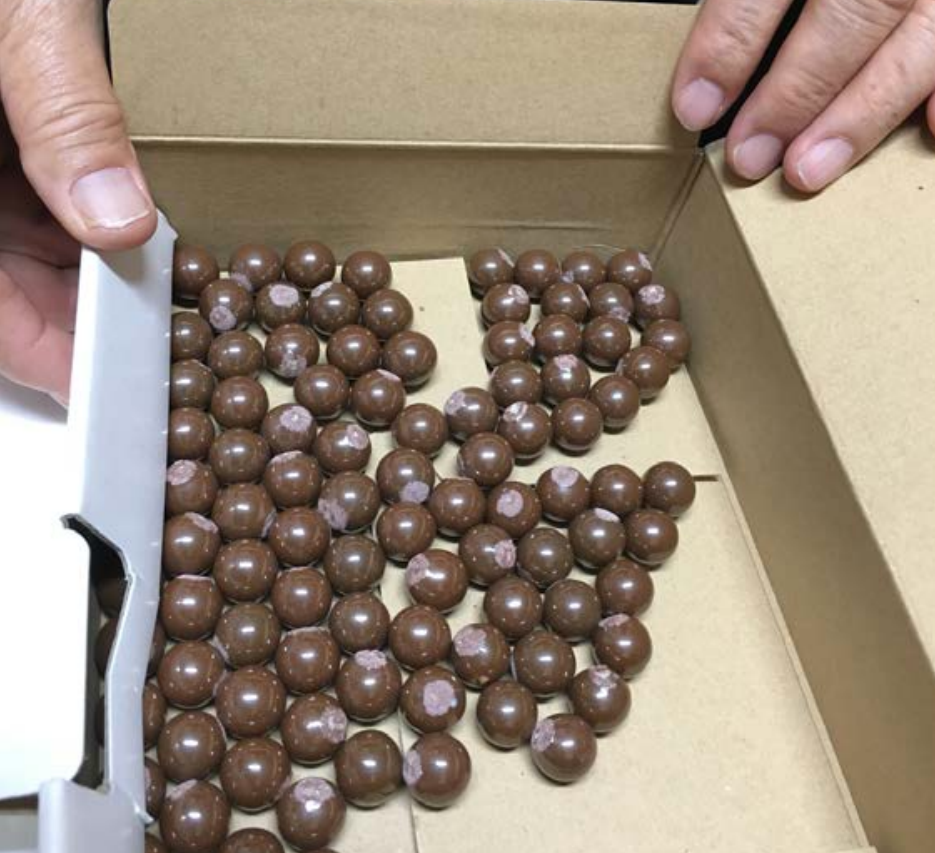}$$
\caption{{\footnotesize Unwanted effect of scale and the device studied here to cure it}}
\label{figone}
\end{center}
\end{figure}

Water pollution is a serious problem everywhere in the world. As a kind of water pollution, scale (hard grayish-white substance) formation often leads to reduced efficiency or severe damage on  cleaning systems or in pipes or other aqueous closed cycle like cooling systems in industrial plants.
 A concrete examples is shown on the left and center in figure \ref{figone}.
 
\begin{figure}[htbp]
\begin{center}
$$\includegraphics[width=5cm]{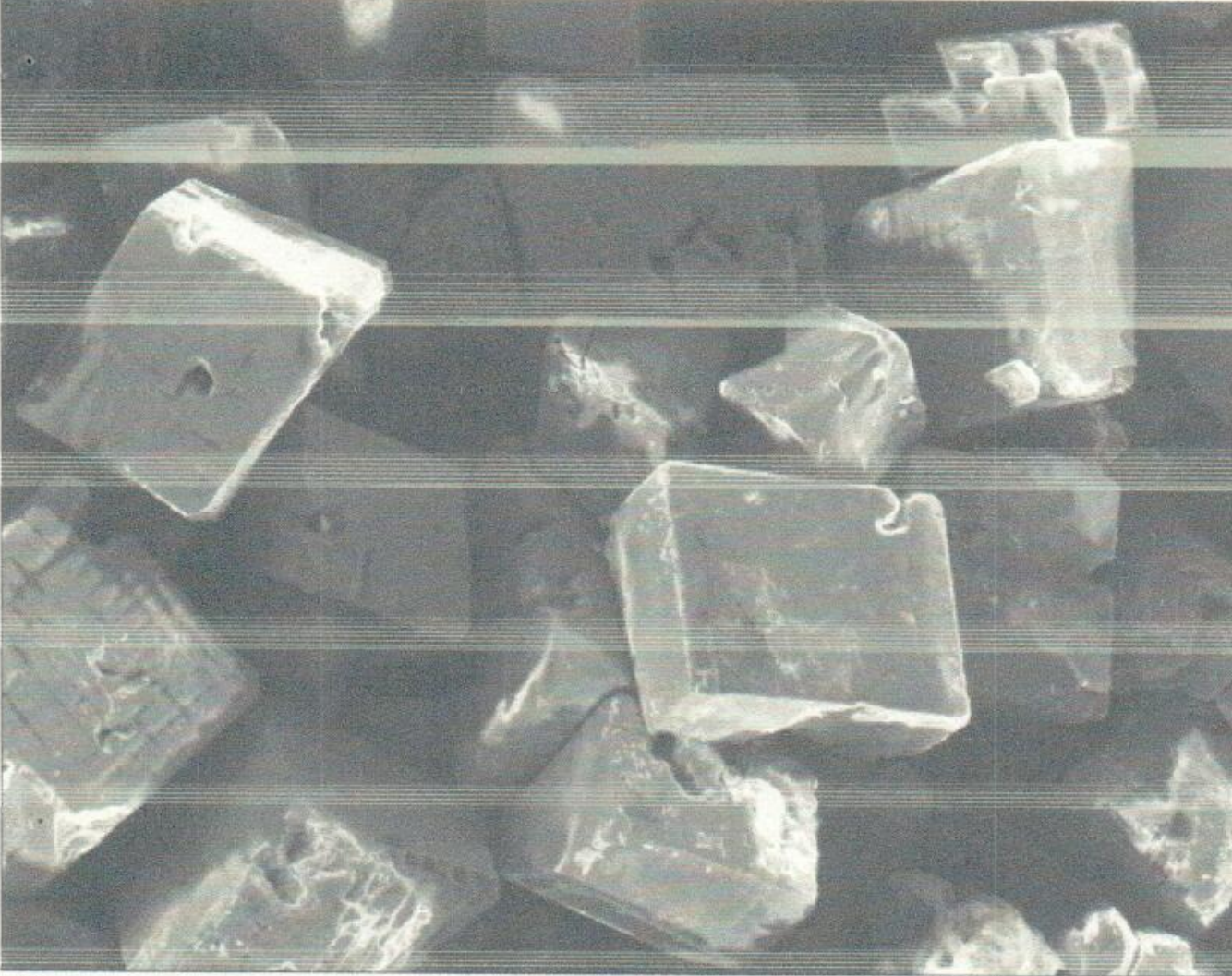} 
~~\includegraphics[width=5cm]{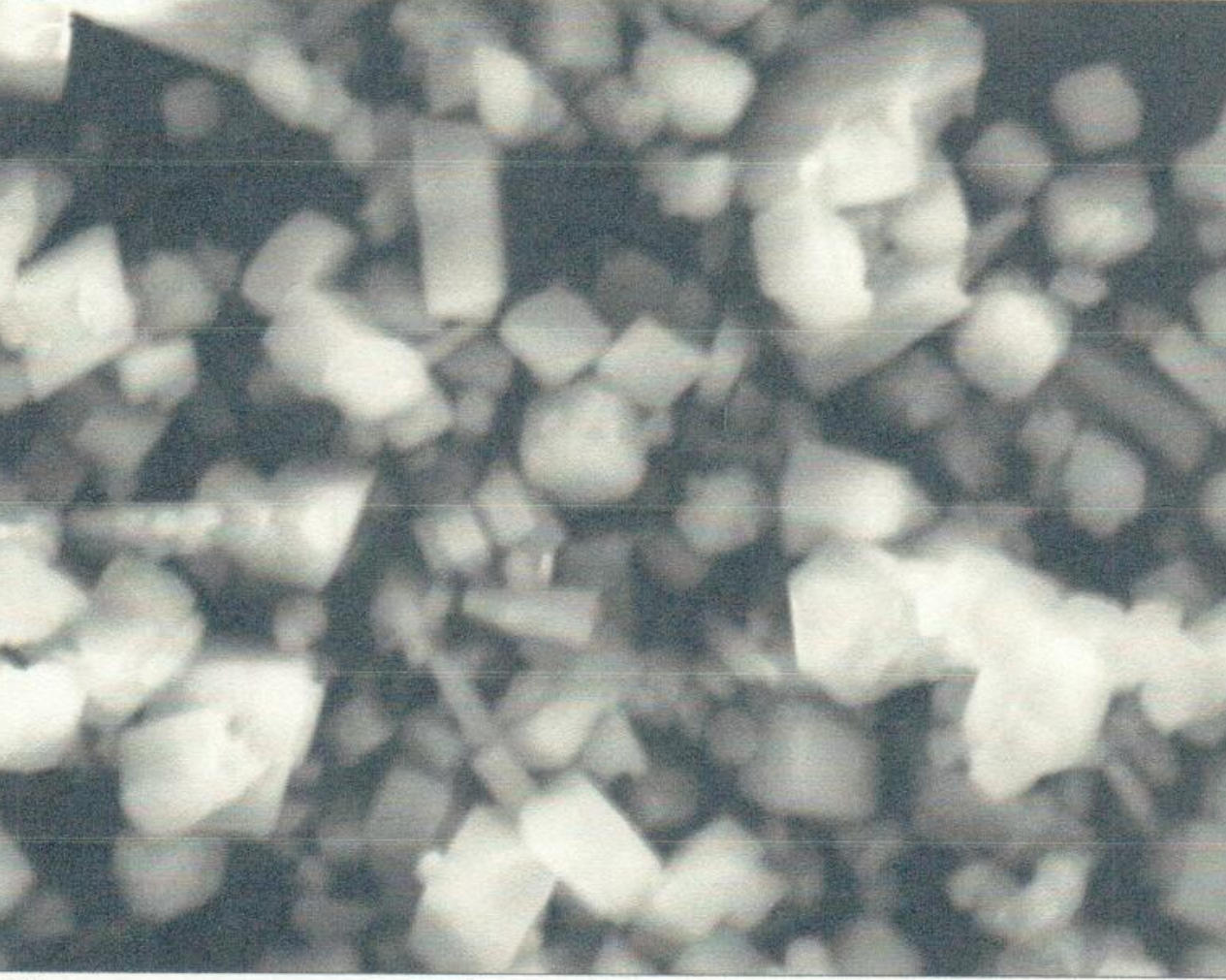}$$
\caption{{\footnotesize Left : \texttt{CaCO}$_3$ crystals in natural water. Right : same after 24h in contact with the spheres of figure \ref{figone}.}}
\label{fig1b}
\end{center}
\end{figure}

Therefore, various anti-scale treatments have been developed for a long time \cite{donaldson,duffy,higashitani,katrane,lopez,mahmoud,sohali,zheng,baker,barrett,gabrielli,higashitaniIseri,higashitaniOshitani,wang,amjd}. One of them is a tight aggregate of ceramic spheres (right of Figure \ref{figone}). If it is immersed  in  natural water, it is known to prevent scale formation \cite{ikawa}. The diameter of the ceramic balls is around 1cm, their surface is covered with inorganic oxides, whose main ingredients are \texttt{SiO}$_2$ and \texttt{Fe}. These produce free electrons by the reaction 
\texttt{Fe}$\to$\texttt{Fe}$^{++}$\texttt{+2e}; in turn the electrons are captured by the hydrogen ions in water \texttt{2H}$^+$\texttt{+2e}$\to$\texttt{H}$_2$ thus leaving a very thin layer of \texttt{OH}$^-$ very near the surface of the balls.
\\\\
It does not to change the water composition contrary to chemical treatments but it changes the size of the calcite  aggregates, as shown on the right in figure \ref{fig1b}.
However, theoretical clarification of their efficiency seems to be open and needs to be worked out. 

\subsection{The Process and the Approach}
What is the cause of scale formation?   Calcite  particles form a crystal by a phase-change phenomenon. Scale is a more stable state for the crystal particles of calcium carbonate (\texttt{CaCO}$_3$, calcite being its most stable crystal form) obtained  by releasing their surface tension energy by adhering to the surface of another material. Experimental measurements reported in \cite{kawarada} have shown that the ceramic spheres  in water generate an electric field near their surface which has an effect on electrolytic solutions such as water containing \texttt{Ca}$^{++}$ and \texttt{CO}$_3^{--}$ ions: the layer of \texttt{OH}$^-$ attracts \texttt{CO}$_3^{--}$ ions and hence change the crystal structure of calcite;  crystal particles passing near the surface of a ceramic sphere are exposed to an electric field and store this additional \emph{polarization} energy which changes the crystal structure making scale formation less favourable energetically.   
 
So our investigations focus first on the electric field  and next on the polarization energy and more precisely on the surface tension energy. We call this model, the \emph{ contact model}. 

 Finally, once it is explained that near the ceramic balls the slate-capable calcite concentration diminishes, it is important to investigate the efficiency of various arrangements of ceramic spheres. This is done by using a macroscopic model and an adhoc boundary condition for the calcite concentration. This last section is mostly numerical. 

\subsection{Plan}
Thus this paper  contributes to the clarification of the mechanism of such device to prevent scale formation in natural water.

In section \ref{sec2} of this article, we derive a Poisson-Boltzmann equation to formulate this phenomenon.
The Poisson-Boltzmann equation is solved numerically for appropriate physical constants and then the electric field on the surface of ceramic balls is obtained. By using it, the polarization energy generated in particle is computed.  

In section \ref{sec3}, it is shown that the polarization energy gives rise to an extraordinary 
suppression of nucleation rate.  This means a substantial  
decrease of nuclei which contribute to formation of the scale. 

In section \ref{sec4}, a framework of crystal lattice model to describe a structure of particle is introduced to compute the surface tension energy. Deviation of the surface tension energy based on the polarization energy is computed by using crystal lattice model. This means the decrease of the surface tension energy which contributes to reduction for formation of the scale.

 In section \ref{sec5}, the distribution of polarized nuclei carried by the water flow through an array of ceramic balls in a pipe is simulated numerically to investigate the behavior of polarized nuclei. The numerical results indicate important information on the best way to install an array of ceramic balls in a pipe.  In appendix, the growth of particle with the polarization is discussed by use of time-dependent Ginzburg-Landau model. Under the assumption that the radius of particle is sphere, it is shown that the time-asymptotic radius of particle is comparable to the radii of the crystals observed by using the standard  \texttt{(NH}$_4$)$_2$\texttt{CO}$_3$ \emph{Gas Diffusion Method} \cite{dietzsh}.  This proves the validity of the contact model in a method. 
    
\section{Electro-chemical phenomena near the surface of a ceramic sphere in water} \label{sec2}
\subsection{Generalities}
It is estimated that, in water, the surface of a ceramic sphere is highly charged.  As the ceramic surface is covered with inorganic oxide, it supplies hydroxyl groups near the surface. An hydroxyl group is charged positively or negatively according to the \texttt{[pH]}. Normally the surface of a ceramic sphere is charged  negatively. Therefore \texttt{Ca}$^{++}$ ions are attracted near the surface and form a cloud.  Similarly \texttt{CO}$_3^{--}$ ions are kept away from the surface. Negative charges on the surface of the ceramic sphere and positive clouds of \texttt{Ca}$^{++}$ ions combine to form a diffused electric double layer, which brings about an electric potential near the surface of the ceramic spheres in water \cite{kitahara,derjaguin}. 

\subsection{Modeling}
Let ${\bf x}\mapsto\phi({\bf x})$ be the electric potential defined at each point ${\bf x}$ outside a   disk in ${\bf R}^2$ or sphere in ${\bf R}^3$.

In natural water the concentration of  \texttt{Ca}$^{++}$ and \texttt{CO}$^ {--}_3$ are uniform and equal to the same constant $C_0$. Throughout the paper $C_0=0.017$\texttt{mol m}$^{-3}$.

In the presence of ceramic balls, because of the electric field, the concentrations $C_1$ o f  \texttt{Ca}$^{++}$ and  $C_2$ of  \texttt{CO}$^ {--}_3$ are no longer equal nor constant; however they are related by $C_1C_2=C_0^2$ due to conservation of flux  (see \cite{komiyama}).

Then the charge density ${\bf x}\mapsto\rho({\bf x})$ made by $C_1$ and $C_2$ is (see \cite{kitahara,derjaguin})
\begin{equation}
      \rho={\bar z} F( C_1 - C_2)
\end{equation}
where ${\bar z}$ is the number of electric charges of the ion, ${\bar z}=2$ for \texttt{Ca}$^{++}$ and the Faraday constant $F$=96485\texttt{(C/mol)} .

Conservation of flux requires (see \cite{komiyama}) $C_1=C_0 e^{{b}\phi}$ and $C_2=C_0 e^{-{b}\phi}$ with ${b}={\bar z} F / (R T)$, where $R =8.3145$\texttt{(J/mol/K)} denotes the perfect gas constant and $T$ is the absolute temperature. When $T=293$ then $b=79.211$. Hence
\begin{equation}\label{rho}
      \rho=2 {\bar z} F C_0\sinh({b}\phi) 
\end{equation}
Furthermore $\phi$ satisfies the Poisson-Boltzmann equation,
\begin{equation}\label{P}
     \Delta \phi = \frac{\rho}{\epsilon_\gamma \epsilon_0} 
\end{equation}
where $\Delta$ is the Laplace operator,  
$\varepsilon_{\gamma}$ the relative permittivity of water and $\varepsilon_{0}$ the permittivity of vacuum:
\begin{eqnarray}\label{5}&& 
\epsilon_\gamma=78.5, ~~~\epsilon_0=8.8542 \cdot  10^{-12}~\hbox{\texttt{(C/V\,m)}}. 
\end{eqnarray}

\subsection{Boundary conditions}
The value of $\phi$ far away from the ball or disk is $\phi = 0$. The value, $\phi_1$, of $\phi$ on the boundary of the ball or disk depends on the matter of inorganic oxides which composes the surface of the ceramic sphere or disk. This is formulated as,
\begin{equation}
   \phi_1 = 
    -0.0591\texttt{[pH]}-0.0295\log[P(\texttt{H}_2)]. 
\end{equation}
This equation is derived from Nernst equation \cite{murase} (equation (20-2)). $P(\texttt{H}_2)$ is the partial pressure of $\texttt{H}_2$, the value of which is taken as 1 atmosphere.   
The present model is intended for $C_0$ in the range $O(0.01)$ mol m$^{-3}$.  

\subsection{Numerical analysis of the Poisson-Boltzman Problem}
We assume spherical or cylindrical symmetry of the domain.  

The domain defined by
 $r:=|{\bf x}|\in (R_b,+\infty)$  and typically $R_b=0.01$.
 With spherical ($d=2$) or cylindrical ($d=1$) symmetry of (\ref{rho})  reduces to  
 \begin{eqnarray*}& \displaystyle
 -\frac1{r^d}\partial_r(r^d\partial_r\phi) + \tilde A\, C_0 \sinh({b}\phi)=0,~~& 
\phi(R_b)=\phi_1,~\phi(R_\infty)=0,
\end{eqnarray*}
with $\tilde A= 2 {\bar z} F /(\epsilon_\gamma\epsilon_0)$.  Note that $d=0$  corresponds to a planar boundary.
\\
Based on the above,
$\tilde A=5.5527\cdot  10^{14}$, $b=79.211$.
The problem is now rescaled by changing the unknown to
$\psi={{b}}\phi,$
then  leading to
\begin{eqnarray}\label{system}&\displaystyle
-\partial_r(r^d\partial_r\psi) +  r^d \tilde A C_0b\sinh(\psi)=0,~ &
\psi(R_b)= b\phi_1,~~\psi(R_\infty)=0,~~~~~~~~~~ 
\end{eqnarray}
Curvature effects being negligible we can take d=0 and consider the equation:
\begin{equation}\label{getpsipp}
-\psi''+\mu\sinh\psi=0,~~\psi(R_b)= b\phi_1,~~\psi(R_\infty)=0
\end{equation}
where $\mu=\tilde A C_0 b$ .  Multiplied by $\psi'$ and integrated the differential equation gives:
\[
\frac12{\psi'}^2 = \mu\cosh\psi + constant
\]
As both $\psi$ and $\psi'$ tend to zero as $r\to R_\infty$, the constant must be equal to $-\mu$. Hence
\begin{equation}\label{getE}
\psi'=-\sqrt{2\mu(\cosh\psi-1)} \hbox{ leading to } \psi'(R_b)=-\sqrt{2\mu(\cosh(b\phi_1)-1)}.
\end{equation}
This formula allows the computation the electric field at the surface of the ball,
\begin{eqnarray}\label{elec}
E&=&-\frac{\partial\phi}{\partial r}(R_b)=-\frac1b\psi'(R_b) =\sqrt{\frac{2\tilde A C_0}b}\sqrt{\cosh(-4.6813\texttt{[pH]})}
\cr&=&0.3744\cdot 10^7\sqrt{C_0\cosh(4.6813\texttt{[pH]})}.
\end{eqnarray}
Variations of $E$ with respect to \texttt{[pH]} and $C_0$ are easy to compute by (\ref{elec}). Some values are given in Tables \ref{pH} and \ref{C0} for $b=79.211$.
\begin{table}[htp]
\caption{{\footnotesize Variation of $E$ with the \texttt{[pH]} when $C_0=0.017$.}}
\begin{center}
\begin{tabular}{|c|c c c c c|}
\hline
\texttt{[pH]}  & 2 & 5 & 7 & 9 & 11\cr
$E$ & $8.179\cdot 10^8$ & $9.169\cdot 10^{11}$ & $9.894\cdot 10^{13}$ & $1.068\cdot 10^{16}$
& $1.151\cdot 10^{18}$\cr
\hline
\end{tabular}
\end{center}
\label{pH}
\end{table}
\begin{table}[htp]
\caption{{\footnotesize Variation of $E$ with $C_0$ when  \texttt{[pH]}=7.}}
\begin{center}
\begin{tabular}{|c|c c c c|}
\hline
$C_0$  & 0.001 & 0.01 & 0.05 & 0.1 \cr
$E$ & $1.984\cdot 10^{8}$  & $2.400\cdot 10^{13}$  & $1.697\cdot 10^{14}$  & $2.400\cdot 10^{14}$ \cr
\hline
\end{tabular}
\end{center}
\label{C0}
\end{table}%
\paragraph{Discussion}
By (\ref{getpsipp}), $\phi''(R_b)= \frac\mu b \sinh(b\phi_1)= -5.5527\cdot 10^{14} C_0\sinh(4.6813\texttt{[pH]})=-0.8033\cdot 10^{27}$.
A calcite cristal being roughly 5 angstrom wide, the variation of the electric field on that distance being proportional to $\phi''$, it is expected to vary enormously. What is a reasonable value of $E$ then?  A numerical solution of (\ref{system}) is reported in the appendix. Accurate results require an impossibly fine grid, hence the under resolved solution provides a lower value for $E$, leading to 
\[
10^7<E<10^{13}.
\]
Experimental measurements in \cite{murase} confirms this analysis.

In the following sections we will use $E\sim 10^{10}$.

\section{Extraordinary Suppression of nucleation rate based on the contact model} \label{sec3}
\subsection{Activation energy of calcium carbonate particle}

Here we suppose the calcite particles are spherical with radius $r$.  
 Let $\delta G$ denote the difference of Gibbs free energy between an electrolytic solution composed of Ca${}^{++}$, CO${}_3^{--}$ in water  and  one with coagulated calcium carbonate in water.. This  change depends on the ions concentration; furthermore $\delta G$ is  negative for solidification found usually in city water (See section \ref{4.3})

Let $\gamma$ be surface tension energy between calcite and water. Then total free energy change $\delta g$ for coagulated. calcium carbonate carbonate is
\begin{eqnarray}
\delta g = 4\pi \gamma r^2 + \frac{4\pi}{3}  r^3 \delta G.
\end{eqnarray}
As the particle grows, $\delta g$ increases until the radius reaches
\begin{eqnarray}\label{ten}
	r^* =- 2\frac\gamma{\delta G}
\end{eqnarray}
Namely, $\delta g$ attains maximum value  
\begin{eqnarray}\label{eleven}
                \delta g^* =
		\frac{16\pi}{3} \frac{\gamma^3}{(\delta G)^2} 
\end{eqnarray}
at $r = r^*$, which is the energy required for the particle to become a particle \cite{brophy}.
This is activation energy.

\subsection{Study of the nucleation caused by the electric field on the surface of ceramic sphere}

Nucleus is the original form to grow up into crystal particle. Nucleation  means that some among coagulated particles with $\delta g^*$ small enough will become isolated nucleus 

 The number of nucleus generated by nucleation per unit time is defined by 
\begin{eqnarray}
I= I_0\exp{\left(-\frac{\delta g^*}{kT}\right)}.
\end{eqnarray}
where $I_0$ is total number of coagulated particles.   Nucleation rate is defined by $I/I_0$ (see \cite{lamer}, \cite{doi}). 
In this section we try to compute the effect on $I$ of  a polarization energy D added  to $\delta g^*$. 

The electric field $E$ at the surface of the ceramic balls brings about a polarization energy density \cite{nakayama} in calcite particles near the ceramic balls
\begin{equation}\label{polar}
	D=\frac{\varepsilon^c_{\gamma} \varepsilon_{0}}2 E^2 \; (\hbox{J/m}^3).
\end{equation}
with $\epsilon^c_\gamma=9.1$ for calcite.
Adding $D$ to the activation energy $\delta g^*$ leads to a new activation energy, which is denoted by $\delta \tilde{g^*}$,
\begin{equation}
              \delta\tilde{g^*} = \delta g^* +\frac{4\pi}{3} {r^*}^3 D
\end{equation}

Some among coagulated particles with $\delta g^*$ change to be nucleus, the number of which is defined by $I = I_0 \exp( ­ \frac{-\delta g^*}{kT})$.
Where $I_0$ is total number of coagulated particles. This is by defiinition the  nucleation rate.:  $\tilde{I}=I_0 \exp{\left(-\frac{\delta\tilde g^*}{kT}\right)}$. 

 In order to verify the effect brought about by the perturbed energy to nucleation rate, we study the ratio $I^*/I$: 
\begin{equation}\label{ii}
\frac{\tilde I}{I}= \exp{\left(-\frac{\delta\tilde g^*-\delta g^*}{kT}\right)}=e^{-q}
~~~~\hbox{ with }  q =\frac{4\pi}{3} {r^*}^3 \frac {D}{kT}.
\end{equation}

\subsection{Suppression of nucleation rate} \label{4.3}

Following \cite{chang} \cite{giese} \cite{murase}, experimental values are

\begin{description}
\item[-]  Surface tension (free) energy: $\gamma=57 \cdot  10^{-3}$ (J/m$^2$). 
\item[-]   $\delta G_0 = - 47.4251$(kJ/mol) =- 1.284 10$^9$(J/m$^3$) .
\item[-] $ \delta G =-\delta G_0 - RT\log{C_0^2}$
\end{description}
Note that $\delta G_0 \le \delta G$. 
Therefore, by (\ref{ten}), $r^* > 0.8903\cdot 10^{-10}$ and by (\ref{polar}) and (\ref{ii}) and the computation of the previous paragraph showing that $E>10^{10}$,
\begin{equation}\label{q}
q > 3.072\cdot 10^{-20} E^2>3.072  ~\Rightarrow~\frac{I^*}I=\exp(-q) < 0.046 .
\end{equation}
Therefore the number of nuclei which contributes to scale formation becomes considerably less; an extraordinary suppression of nucleation rate occurs due to the contact model.

\newpage

\section{Framework of Crystal Lattice Model}\label{sec4}

Another contribution to the polarization energy for scale prevention is the decrease of surface tension energy of the particle in water due to the structural change of calcite caused by  the perturbed energy .
\\
In order to investigate this phenomenon, we adopt the crystal lattice model developed by S. Ono \cite{ono} and apply it to the crystal structure of calcite. Considerably good agreement between theory and experiments is reported in his book.

\subsection{Helmholtz free energy defined  for a crystal}

\begin{figure}[htbp]
\begin{center}
\includegraphics[width=0.4\textwidth]{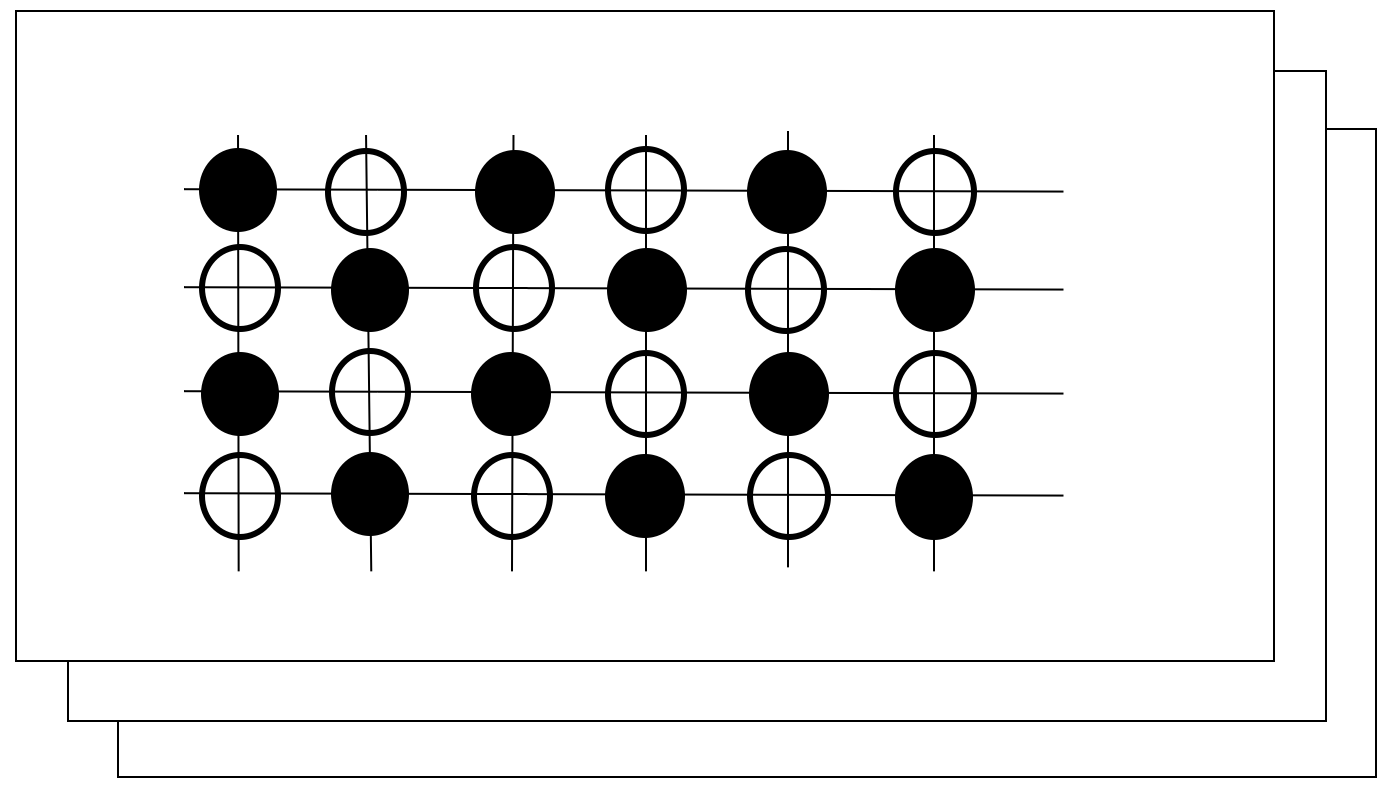}~~
\includegraphics[width=0.4\textwidth]{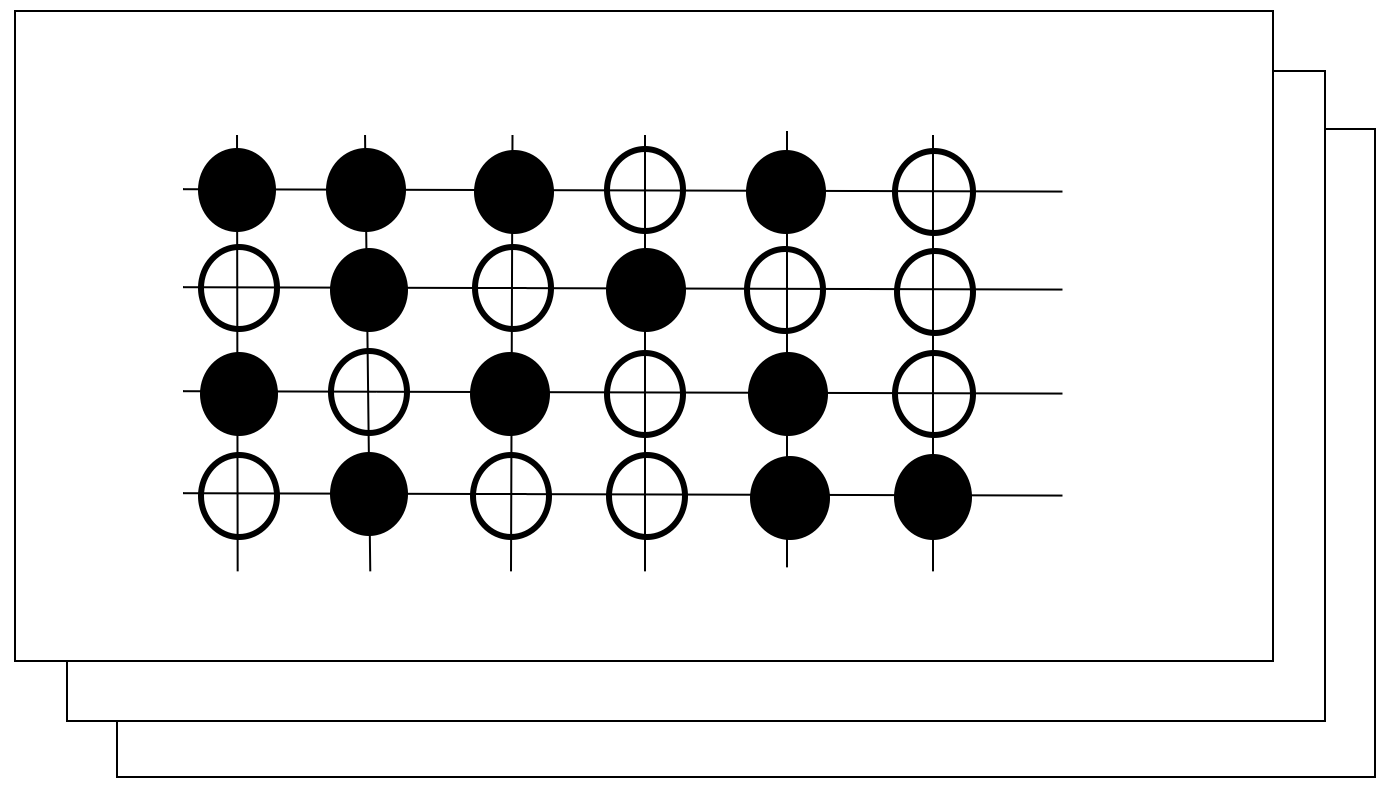}
\caption{{\footnotesize Multi-layers crystal structure at zero Kelvin is periodic (left) while somewhat random at ambiant temperature (right).}}
\label{fig:CrystalStructure}
\end{center}
\end{figure}

The 3-dimensional crystal lattice model assumes that the crystal is composed of molecules arrayed on lattice points (See Figure \ref{fig:CrystalStructure}).
The interaction between molecules is restricted to those which are side by side.

Let  $S$ be the entropy. A state with small entropy is a special state which occurs under some restrictions. If these are removed,  many more possible states will be observed with greater entropy. 
\\
The crystal tends to stay at a state of minimum energy at temperature 0K.  Then all molecules try to fill the empty positions (the white holes of Figure \label{fig4} on the left) and crystal becomes a complete crystal; let $X_0$ be the number of such molecules. In this situation the number of configuration is $W=1$.
\\
If the temperature rises, some of molecules leave the black holes and try to occupy other white holes. Let $X_1$ be the the number of such molecules . The total number of all different configurations is \cite{reif}
\[
W(X,X_1)=\frac{X_0!}{X_1! \cdot (X_0-X_1)!}.
\]
The Boltzmann principle says that 
\begin{eqnarray}
	S = k \log{W(X_0, X_1)}.
\end{eqnarray}
Using  Sterling's formula, an approximate formula is obtained
\begin{eqnarray}
S=-X_0k\left\{ Z \log{Z} + (1-Z) \log{(1-Z)}\right\}
\end{eqnarray}
where $Z= X_1/ X_0$ and $k$ is Boltzmann constant and layers are supposed uniform.

On the other hand, let the number of adjacent molecules depending on the lattice structure be $\bar\alpha$, then an internal free energy $U$ of this system is 
\begin{eqnarray}
	U = -\frac e2\bar\alpha X_0  Z^2.
\end{eqnarray}
where $-e$ is the potential energy between adjacent molecules experimentally measured to be (see \cite{archer})  $e=4\cdot 10^9/\rho$ \texttt{(J)}. 
Therefore Helmholtz free energy $F$ for this arrangement is  
\begin{eqnarray}
	F = U - TS.
\end{eqnarray}

\paragraph{Notations}
 It should be noted that the crystal is composed by many layers.
Here we have considered the free surface of a plane layer of crystal parallel to the other lattice layers.  To analyse a stack of lattice layers we label them from $1$ to $m$.

$A$ will stand for the area of each layer; $n$ will denote the umber of lattice points per unit area on each layer, and
$n_i$  the number of molecules included in a unit area  of the i-th layer.
The total number of molecules is denoted $N$.
Then  $x_i = n_i/N$ is the concentration of molecules per unit surface in the i-th layer.

Finally $h$ will denote the polarization energy per molecule and  $\rho$ the density of calcium carbonate, namely $\rho=1.625\cdot 10^{28}$ m$^{-3}$

\subsection{Multi-layered Crystal}
In order to compute the internal free energy of a crystal made of a stack of 2-dimensional layers, it is noted that the number of neighboring molecules of one molecule is 6 for the same layer and 3 for each neighboring layer. 

The particle pairing number between the $i$-th layer and its adjacent neigbhour $i+1$ is $3nA x_{i+1} x_i$ -- and similarly for $i-1$ -- while the particle pairing with the ith layer is $6nA x_i^2$.  Removing the pairs counted twice in this enumeration, comes to
\begin{eqnarray}
U &=&- 3e\,nA [x_1(x_1+ \frac12 x_2)
      + \sum_{i=2}^{m} x_i(x_i + \frac12(x_{i-1} + x_{i+1}))] \label{eqn:U} \\
S &=&-  n A k\sum_{i=1}^{m} (x_i \ln x_i +(1-x_i) \ln(1-x_i))
\end{eqnarray}

The total energy, $\bar{\bar F}$, is the sum of Helmholtz' free energy  ${{\bar F}} = U - TS$ with the polarisation energy $h\,x_1$ due to the contact model where  $h=D/\rho$ and $D$ is defined in (\ref{polar}) (see \cite{ono}):
\begin{eqnarray}
 \bar{\bar F} &=& n\, A \, h x_1 - 3e\,nA[x_1(x_1+ \frac12 x_2)
      + \sum_{i=2}^{m} x_i(x_i + \frac12(x_{i-1} + x_{i+1})) \nonumber \\
	 &+& 
   T n A k\sum_{i=1}^{m} (x_i \ln x_i +(1-x_i) \ln(1-x_i))~~~
\end{eqnarray}

\subsection{ Minimization problem}

The stable state of the system is obtained by minimizing  ${\bar{\bar F}}$ under the constraint that the chemical potential energy $n\,A\,\eta\sum_{i=1}^m x_i$ is constant; $\eta$ is the chemical potential per particle :
\begin{equation}\label{eta}
\eta=-12 z e + k T \ln\frac z{1-z}.
\end{equation}
 With $a=k T/e=\frac{kT}e=0.01575$. and $\beta=h/e$, the optimality conditions are 
\begin{eqnarray} 
\label{euler3}
\beta -3 \left(2 x_1 +x_2 \right) +  a \ln{\frac{x_1}{1-x_1}}  &=& -12z+a\ln\frac{z}{1-z}\\ \label{euler1} 
-3 \left(x_1 +2x_2 +x_3 \right)  +  a \ln{\frac{x_2}{1-x_2}} &=& -12z+a\ln\frac{z}{1-z} \\ 
-3 \left(x_{i-1} +2x_i +x_{i+1} \right)  +  a \ln{\frac{x_i}{1-x_i}} &=& -12z+a\ln\frac{z}{1-z} \\ 
  \label{euler2}
-3 \left(x_{m-1} +3x_m \right) +  a \ln{\frac{x_m}{1-x_m}}  &=& -12z+a\ln\frac{z}{1-z}.
\end{eqnarray}
If $\beta=-3z$ the solution is $x_i=z,~i=1\dots m$. \\\\
Let the number of layers be large and $\beta+3z$ be small. Then  let $x_i=z+x'_i$ and $\beta'=\beta+3z$ so that
\begin{eqnarray} 
-\beta' +3 \left(2 x'_1 +x'_2 \right) &=&  a \ln{\frac{1+\frac{x'_1}z}{1-\frac{x'_1}{1-z}}} 
\approx a(\frac{x'_1}z+\frac{x'_1}{1-z}) \\  
3 \left(x'_{i-1} +2x'_i +x'_{i+1} \right)  &=&  a \ln{\frac{1+\frac{x'_i}z}{1-\frac{x'_i}{1-z}}}
\approx a(\frac{x'_i}z+\frac{x'_i}{1-z})\\ 
3 \left(x'_{m-1} +3x'_m \right) &=&  a \ln{\frac{1+\frac{x'_m}z}{1-\frac{x'_m}{1-z}}}\approx a(\frac{x'_m}z+\frac{x'_m}{1-z}).
\end{eqnarray}
Notice that a finite difference approximation of $\displaystyle\alpha u-\frac{d^2 u}{d y^2}=0$ on a uniform grid of size $dy$ is $\alpha u_i-(u_{i-1}-2u_i+u_{i+1}/{\delta y}^2$ . By analogy we see that the above is a finite difference discretization of
\[
-3{\delta y}^2\frac{d^2 u}{d y^2}(y) + [a(\frac1z+\frac1{1-z})-12{\delta y}^2]u(y)=0,~y\in(0,1),~~u(0)=-\frac{\beta'} 3,~~\frac{d u}{d y}(1)=0,
\]
the solution of which is $u=c e^{\alpha y} + b e^{-\alpha y}$ with 
\[
\alpha=\sqrt{\frac a{3{\delta y}^2}(\frac1z+\frac1{1-z})-4},~~c=\frac{\beta'e^{-2\alpha }}{3(1+e^{-2\alpha })}, ~~ b=\frac{\beta'}{3(1+e^{-2\alpha })}.
\]
Thus
\[
x'_i\approx u(ih)\approx \frac{\beta'}3 \exp\left(-i\sqrt{\frac a{3z(1-z)}}\right)
~~\Rightarrow~~x_i\approx z+(\frac\beta 3 +z)\exp\left(-i\sqrt{\frac a{3z(1-z)}}\right).
\]
Let us use this solution to compute by (\ref{eqn:U}) the Helmholtz energy for each layer of the uniform phase (the bulk) . The chemical potential is defined below; the Helmholtz free energy for each layer of the uniform phase (bulk)is defined
\begin{eqnarray}
\tilde F&=&{n A} \cdot [-6z^2 \cdot e + kT \cdot
	\{z \cdot \ln{z} + (1-z)\ln{(1-z)}\}]. \label{eqn:Fu}
\end{eqnarray}
where $z$ is the density of molecule per unit surface, i.e. the limit of $x_m$ for $m$ large.

The term $-6z^2$ in the right hand side is obtained by taking $x_m=x_{m-1}=z$ in the last term of the  sum in the right hand side of (\ref{eqn:U}). 
\\
Let $m \rightarrow \infty$. Then we may regard $\eta$ as the chemical potential \cite{ono} and
\begin{eqnarray}
\eta = \frac{1}{nA} \cdot \frac{d\tilde F}{dz} = -12z \cdot e + kT \cdot \ln{\frac{z}{1-z}}.
\end{eqnarray}


\subsection{Computation of Surface Density and Polarization Energy}
The experimental value of molecular entropy for calcium is 4.54 (J/K/mol) at room temperature, and per particle it is 4.54/$N_A$, where $N_A$ is the Avogadro number \cite{staveley}.  Equating it with the theoretical value gives
\[
\frac{4.54}{N_A}=-k\left(\frac{X_1}{X_0}\ln\frac{X_1}{X_0} + (1-\frac{X_1}{X_0})\ln(1-\frac{X_1}{X_0})\right)
~\Rightarrow~X_1=0.2348 X_0
\]
The number of particles per unit volume being $(X_1/X_0)^{2/3}$,  
\begin{equation}\label{z}
z=(0.2348)^{2/3}=0.3806.
\end{equation}
The polarization energy per particle, $\beta=h/e=a D/(kT\rho)$ is
\begin{equation}\label{beta}
\beta= 0.1007 \cdot 10^{-20}E^2.
\end{equation}

 \subsection{Deviation of surface tension energy for particle under the contact model}
\subsubsection{ Definition of Surface Tension Energy} 
Following  \cite{ono}, surface tension energy is defined by
\[
            \gamma ( x_1, x_2,... ,x_m, z) = \frac{{\bar{\bar F}}- m \tilde F}A - n \eta\sum_{i=1}^{m}(x_i -z) 
\]
The sensitivity of $\gamma$  with respect to $x_i$ and $z$ is
\[
           \delta \gamma = \sum_{i=1}^{m}\frac{\partial\gamma}{\partial x_i}\delta x_i +  \frac{\partial\gamma}{\partial z}\delta z
\]
In view of the definition of ${\bar{\bar F}}$, at equilibrium $\partial \gamma/\partial x_i = 0$ for $i=1,2,\cdots,m$. Hence 
\[
          \delta \gamma
          = - m (-12z \cdot e + kT \cdot \ln{\frac{z}{1-z}})\delta z  - n  \frac{\partial\eta}{\partial z}\sum_{i=1}^{m} (x_i-z) \delta z -n\eta\sum_1^m(-1)
 \]
Here we used (\ref{eta}) to compute $\frac{\partial\eta}{\partial z}$ to obtain
\begin{equation}\label{dgama}
\delta \gamma
          =  -n(-12 e + \frac{k T}{z(1-z)})\sum_1^m(x_i-z)\delta z = -n\frac{\partial\eta}{\partial z}\delta z\sum_1^m(x_i-z)
\end{equation}

\subsubsection{Formula for $\delta\gamma/\gamma$}
Let us notice that an additional energy $h=\displaystyle\frac{\partial\eta}{\partial z}\delta z$ corresponds to the increase of chemical potential $\eta$ for the surface layer; $h$ is related to the polarisation energy defined in (\ref{polar}) by $D=\rho h$.
\\\\
Using only the first layer in (\ref{dgama}) we have $\delta\gamma=-n h  (x_1-z)$ where $n$ is the surface density of particle, related to $\rho$ by $n=\rho^{2/3}$.
Using the value for $\gamma$ given in section \ref{4.3} , we have
\[
\frac{\delta\gamma }{\gamma}= \rho^{2/3}\frac D{\rho\gamma} (x_1-z)
\]
\subsubsection{Computation of $\delta\gamma/\gamma$}
If we set $E=\alpha 10^{10}$ (see section \ref{sec2}),  $\beta=0.1007\alpha^2$ (see (\ref{beta})) then
\[
x_1-z = (\frac\beta 3 +z) \exp(-\sqrt{\frac a{3z(1-z)}})  = 0.2892\alpha^2+0.3278.
\]
It follows that  $\delta\gamma/\gamma=-0.807\alpha^2(\alpha^2+1.134)$.
%
\\\\
It is pointed out in \cite{higashitani} that the perturbation energy on the particle of calcium carbonate is effective  for the prevention of scale formation 
only in a specific range of \texttt{[pH]}, $C_0$ and P(H$_2$).
\\
Note that when $\alpha^2=2.553$, $\delta\gamma/\gamma=-1$ approximately.  Then $\gamma$ becomes $\gamma + \delta \gamma$ which is also $\gamma(1+\delta\gamma/\gamma)=0$; hence there is no surface tension energy, or at the least heavily reduced.

\section{Numerical Simulations of Calcite Carried by the Flow through an Array of Ceramic Spheres}\label{sec5}
We have seen above that calcite particles undergo a change very near the ceramic spheres due to polarisation.
As this change is very fast, vorticity in the flow  near the surface of ceramic spheres increases the number of particles affected by polarisation.  
The water flow being at moderate Reynolds number, it is well modelled by the Navier-Stokes equation (\ref{nseq}) with zero velocity on the balls. The concentration of calcite is modelled by a convection diffusion equation, but the boundary condition needs to reflect the fact that the ceramic balls decrease the concentration and more so as the vorticity is large which reflect the fact that upstream face  of the ball is more active than its down stream side. 

The purpose of this section is to measure the macroscopic effect of an array of ceramic balls on a flow of calcite in water.  We make  two groups of simulations; one with an inflow on the right and an outflow on the left in a box-like container. The parameters are the geometrical arrangements of the ceramic balls, the pressure difference between the inlet and the outlet and the size of the container.  

The second group of simulations is an attempt to replicate numerically the experiment of \cite{kawarada}.  It is a cylindric container with a rotating magnet below it which induces a rotation of the balls around the axis of the cylinder. There also there are many geometrical parameters in addition to the roating velocity.

For all simulations, the polarisation efficiency of the ceramic balls is an important last parameter, but as we have no access to it, it must be guessed; hence this section remains phenomenological, giving trends rather than experimentally comparable numbers.

\subsection{ The Partial Differential Equations}
The flow is solution of the Navier-Stokes equations:
\begin{equation}\label{nseq}
 \partial_t u + u\nabla u + \nabla p - \nu \Delta u=0,~~\nabla\cdot u=0 . 
\end{equation}
Here $p$ is the pressure and $u$ is the velocity. The reduced viscosity is fixed at $\nu=1/500$. The velocity is given at $t=0$:  $u=0$; it is also zero on the spheres (disks) and given on the walls of the container except at the inlet and outlet where we prefer to impose a pressure gradient and a zero tangential velocity.
\\\
The water is charged with calcite at concentration $C_0$ initially and at the inlet pipes. Everywhere else we expect the concentration of calcite to be $C(x,t)\leq C_0$; without loss of generality we can take $C_0=1$. The reaction to reduce calcite occurs mainly within a very thin boundary layer around each sphere, modelled by 
\begin{equation}\label{cbdy}
\epsilon\frac{\partial C}{\partial n} = \xi~|\sigma_n|~C
\end{equation}
where $n$ is the normal to the spheres, $\sigma_n$ is the normal stress in the fluid, i.e. $\displaystyle\sigma_n=p n + \nu\frac{\partial u}{\partial n} $ and $|\sigma_n|^2=p^2+\nu^2(\frac{\partial u_s}{\partial n})^2$.  Intuitively a strong pressure will bring the  calcite nearer to the balls and a strong tangential part of the normal stress will slow down the calcite particle forcing them to stay longer near the balls.
The coefficient $\xi$ is a decreasing function of $\beta$ which governs the absorption rate due to magnetic spheres. It depends also on the reaction rate of calcite transformation and to some extend on the time the calcite stays near a ball i.e. on the fluid velocity $u$.
In addition, the equation for $C$ is
\[ 
\partial_t C + u\nabla C - \epsilon \Delta C=0, 
\]
with $C=0$ at the walls, $C=C_0$ at the in-pipe and $\partial C/\partial n=0$ at the exit pipe and (\ref{cbdy}) on the spheres.
The variational formulation is: find $C$ such that for all $\hat C$,
\begin{equation}\label{cconcentr}
\int_\Omega( \hat C(\partial_t C + u\nabla C) + \epsilon \nabla\hat C\cdot\nabla C + \int_{spheres}\xi|\sigma_n|\hat C C
=0
\end{equation}
\begin{remark}
It is difficult to derive theoretically a reliable formula for $\xi$ in terms of $\beta$.  The molecular diffusion $\epsilon$ needs also to be measured but it is small; in all computations below $\epsilon=\nu$.
\end{remark}

\subsection{Flow in a Cavity through an Array of Ceramic balls}
In a parallelepiped shaped container of size $X_l\times Y_l\times Z_l=1\times 0.7\times 1$ (in meters) water flows from the in-pipe on the right to the exit pipe on the left. The diameter of the pipes is $w=0.1$; the center of the inflow pipe is at $(X_l,Y_b+w/2,Z_l/2)$ and for the outflow pipe at $(0,Y_b+w/2,Z_l/2)$. Here $Y_b=T_l/2$. The radius of the spheres is $r=0.012$.  

The number of spheres in the directions $x,y,z$ are respectively $m_a,n_a,o_a$. These are evenly distributed in the parallelepiped of size $x_c,y_c,z_c$ and centered at $X_a,Y_a,Z_a$, so that there are spheres 	on the corners of the parallelepiped.
\begin{figure}[hbt]
\begin{center}
\begin{tikzpicture}[scale=4]
\draw  (0,0) -- (1,0) -- (1,0.36); 
\draw (1,0.4) -- (1,0.7) -- (0,0.7) -- (0, 0.4);
\draw (0,0.36) -- (0,0);
\draw [dotted] (0.3,0.15) -- (0.7,0.15) -- (0.7,0.55) -- (0.3,0.55) -- cycle;
\draw [fill] (0.3,0.15) circle [radius=0.012];
\draw [fill] (0.35,0.15) circle [radius=0.012];
\draw [fill] (0.4,0.15) circle [radius=0.012];
\draw [fill] (0.45,0.15) circle [radius=0.012];
\draw [fill] (0.3,0.2) circle [radius=0.012];
\draw [fill] (0.35,0.2) circle [radius=0.012];
\draw [fill] (0.4,0.2) circle [radius=0.012];
\draw [red,thick,densely dotted, <->] (0,0.35) -- (0.49,0.35) ;
\draw [red,thick,densely dotted,<->] (0.5,0.) -- (0.5,0.34) ;
\draw (0.2,0.39) node [right] {$X_a$};
\draw (0.5,0.16) node [right] {$Y_a$};
\draw [red,thick,densely dotted, <->] (0,0.75) -- (1,0.75) ;
\draw [red,thick,densely dotted,<->] (1.1,0.) -- (1.1,0.7) ;
\draw (0.5,0.8) node [right] {$X_l$};
\draw (1.1,0.35) node [right] {$Y_l$};
\draw [blue] (0.98,0.378) node [right] {{\tiny in}};
\draw [blue] (0.,0.375) node [left] {{\tiny out}};
\draw [red,thick,densely dotted,<->] (0.9,0.) -- (0.9,0.36) ;
\draw (0.88,0.15) node [right] {$Y_b$};
\draw [blue,thick,densely dotted, <->] (0.3,0.57) -- (0.7,0.57) ;
\draw [blue,thick,densely dotted,<->] (0.72,0.15) -- (0.72,0.55) ;
\draw (0.48,0.6) node [right] {$x_c$};
\draw (0.73,0.35) node [right] {$y_c$};

\end{tikzpicture}
\caption{{\footnotesize Sketch of the computational domain: a rectangle $(0,X_l)\times(0,Y_l)$ minus the black disks. These are evenly distributed in a rectangle $(X_a-x_c/2,X_a+x_c/2)\times(Y_a-y_c/2,Y_a+y_c/2)$. Water charged with calcite enters from a hole on the right vertical border and exit by a hole on the left border. These are the two segments $\{X_l\}\times(Y_a, Y_a+w)$ and $\{0\}\times(Y_a, Y_a+w)$}}
\label{sketch}
\end{center}
\end{figure}
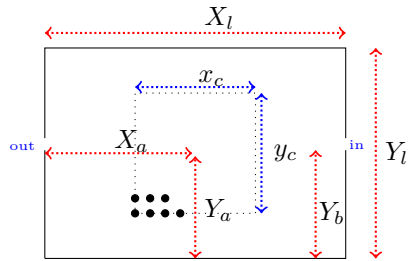
The problem is solved in 2D.  We shall attempt to measure the errors connected with such an approximation at the end of this section.

 A pressure gradient is imposed between the inlet and the outlet $p_{in}-p_{out}=10$; this value is arbitrary but it fixes Reynolds number at a numerical value of a few hundred, compatible with the mesh resolution used here (around 15000 vertices). The computed horizontal velocity due to the pressure gradient depends on the layout of the spheres and is around 7; thus the Reynolds number based on the disk diameter is 70 and 3500 when based on the size of the domain.

\begin{figure}[htbp]
\begin{center}
\includegraphics[width=6cm]{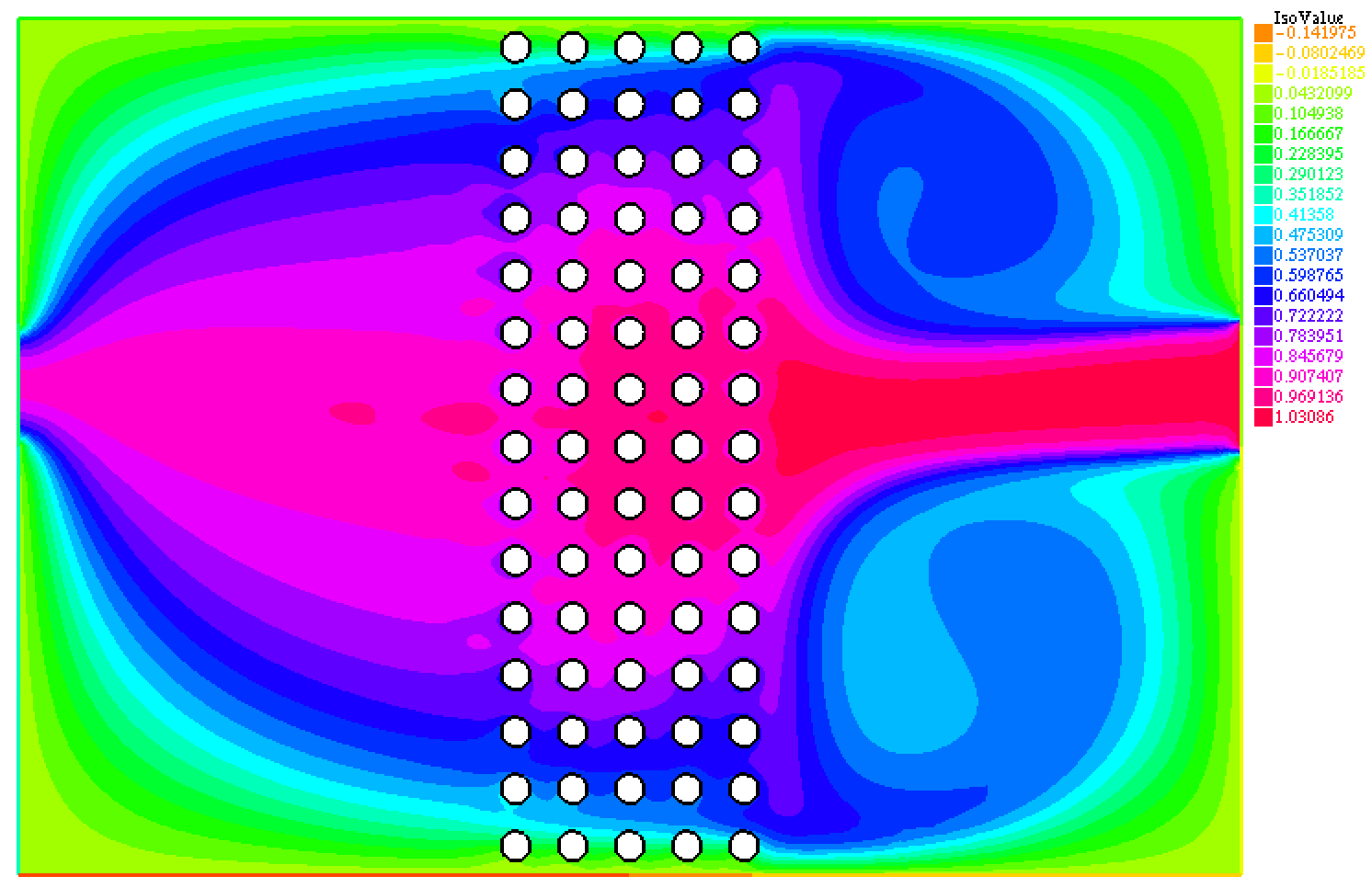}~~
\includegraphics[width=6cm]{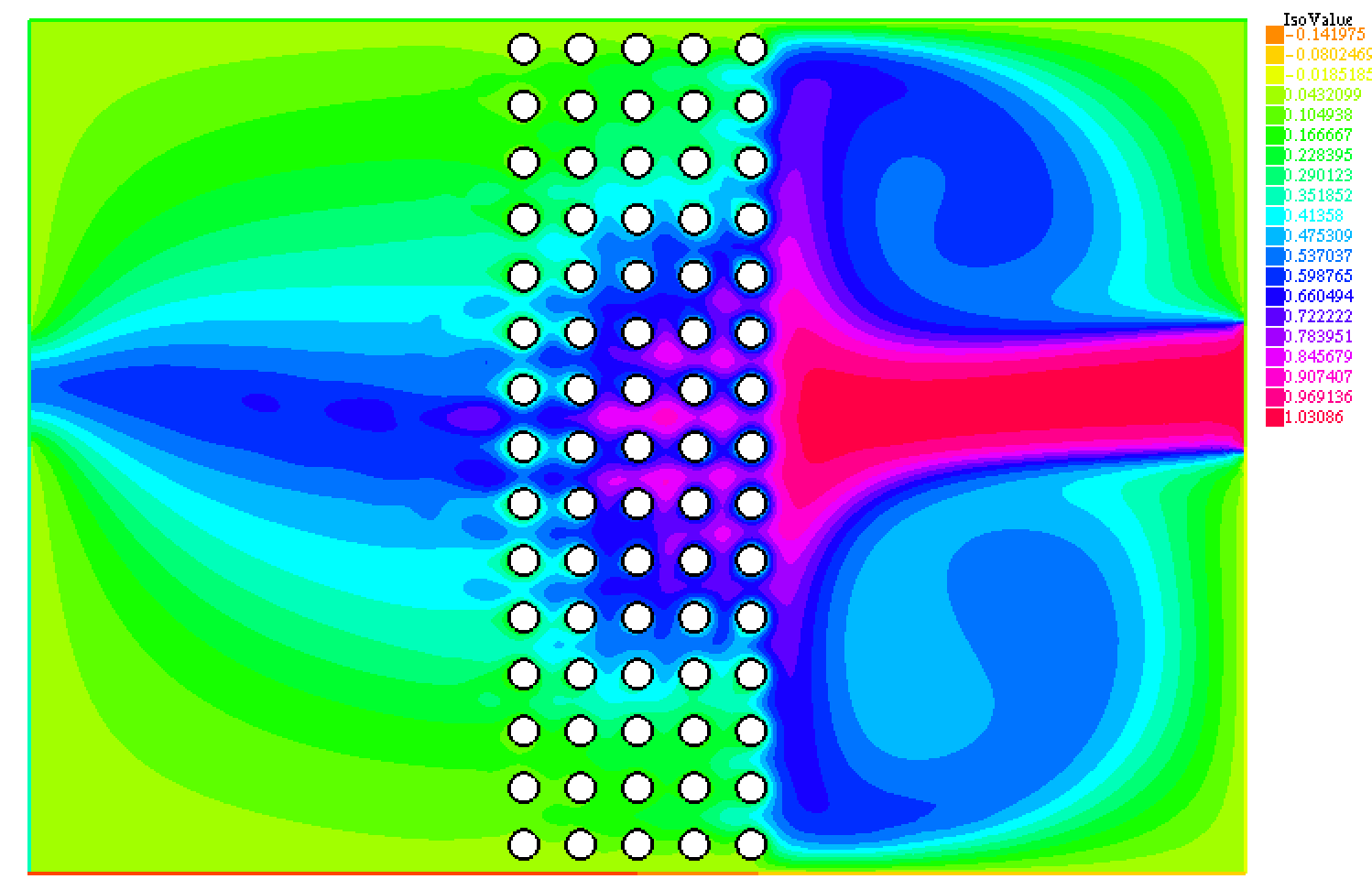}\\
\caption{{\footnotesize {Display of $C$ at T=2.5 after 50 time steps (nearly stationary)  showing the transformation of hard water (red) into soft water(blue). The water runs from right to left. On the right the flows enter saturated with calcite (red) by an aperture. It flows through an array of magnetic disks, looses its calcite and exits through a section slightly above the middle  of the left vertical boundary. Left figure: $\xi=0.02$. Right figure, $\xi=0.2$. }}}
\label{flowconcentration1}
\end{center}
\end{figure}

\begin{figure}[ht]
\begin{center}
	\begin{minipage}[b]{0.45\linewidth}
	\centering
\includegraphics[width=6cm]{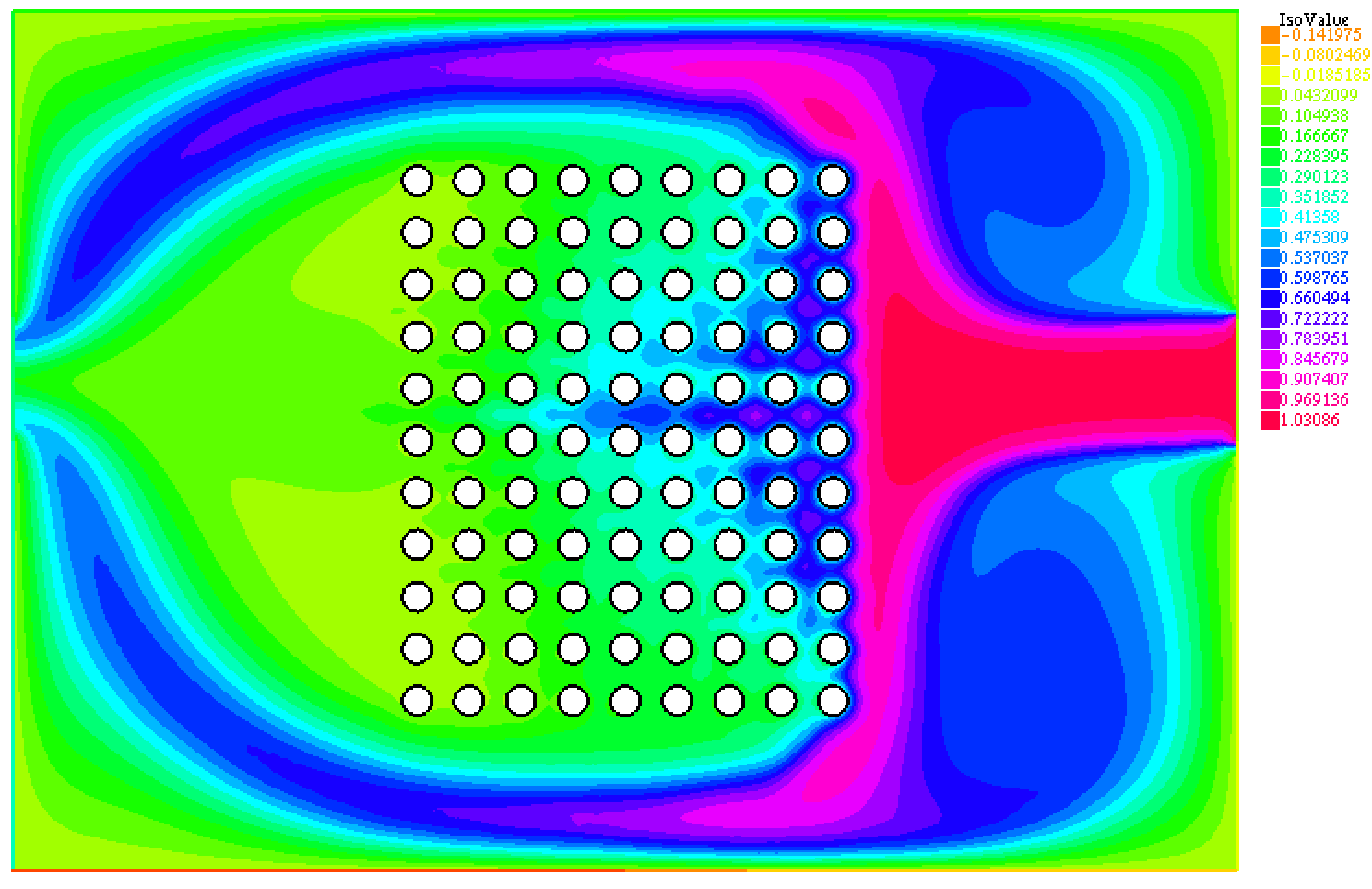}
\caption{{\footnotesize {Display of $C$ at T=2.5 after 50 time steps for $\xi=0.2$ for a different geometrical configuration which proves to be less efficient. It show that the array of spheres should block the pipe fully.}}}
\label{flowconcentration2}
	\end{minipage}
	\hspace{0.7cm}
	\begin{minipage}[b]{0.45\linewidth}
	\centering
\includegraphics[width=4cm]{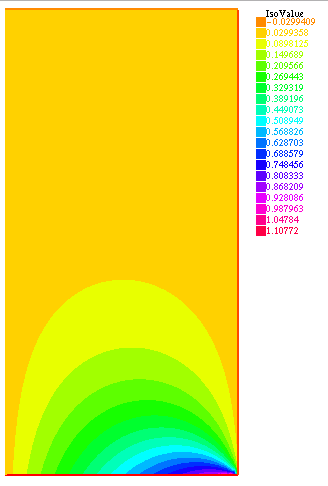}
\caption{\label{axiveloc}{\footnotesize
Angular velocity in a cylinder with a rotating bottom}}
	\end{minipage}
\end{center}
\end{figure}

Figure \ref{flowconcentration1} shows the concentration levels at T=2.5 after 50 time steps for $\xi=0.02$ and $0.2$.  A reduction  is obtained between $C_{in}$ and $C_{out}$ seen in colors changing from red to blue; the reduction is much more intense with $\xi=0.2$.
\\
Figure \ref{flowconcentration2} shows the concentration levels at T=2.5 after 50 time steps for $\xi=0.2$ for a different geometrical configuration.  It is clear that such a design which let the fluid flow through the container without coming into contact with the balls is a bad design.  

The effect of a stronger pressure gradient is obvious and not shown here:  above a certain threshold there is only a partial transformation of the calcite, unless the number of rows of balls is increased.  Below the threshold the calcite lingers unnecessarily long near the balls.
\subsection{Rotating Flow in a Cylinder with Ceramic Balls}
In an experiment reported in \cite{kawarada}, a rotating magnetic agitator below the container induces a rotational flow with an added magnetic agent \bb{ is it so} in a circular cylinder where sits, near the bottom, several layers of magnetically coated balls which do not rotate with the flow.

For the simulation it is easier to assume that it is the bottom of the cylinder which is rotating while the balls and the vertical boundary of the cylinder stay put.

If the balls are ignored, the solution of the Navier-Stokes equations in a cylinder with prescribed angular velocity on the bottom boundary is, in cylindrical coordinate, $u_r=u_z=0$, $u_\theta$ solution in a rectangle of
\[
-\frac1r\partial_r(r\partial_r u_\theta) + \frac{u_\theta}{r^2}-\partial_{zz}u_\theta=0
\]
with $u_\theta=\omega r$ at the bottom of the cylinder and  zero on the side.

The solution is shown on figure \ref{axiveloc} when $\omega=1$.

Next we zoom-in on a row of fixed balls in such a flow. We assume that at one point in the cylinder we can reproduce $u_\theta$ by a pressure gradient according to the formula $\frac1r\partial_\theta p=u_\theta$ and then solve the Navier-Stokes equations locally around the spheres with such a pressure gradient to see what happens to the calcite.  We begin with a 2D simulation and then end with a small 3D configuration.
\subsubsection{2D simulation}
\begin{figure}[htbp]
\begin{center}
\includegraphics[width=12cm]{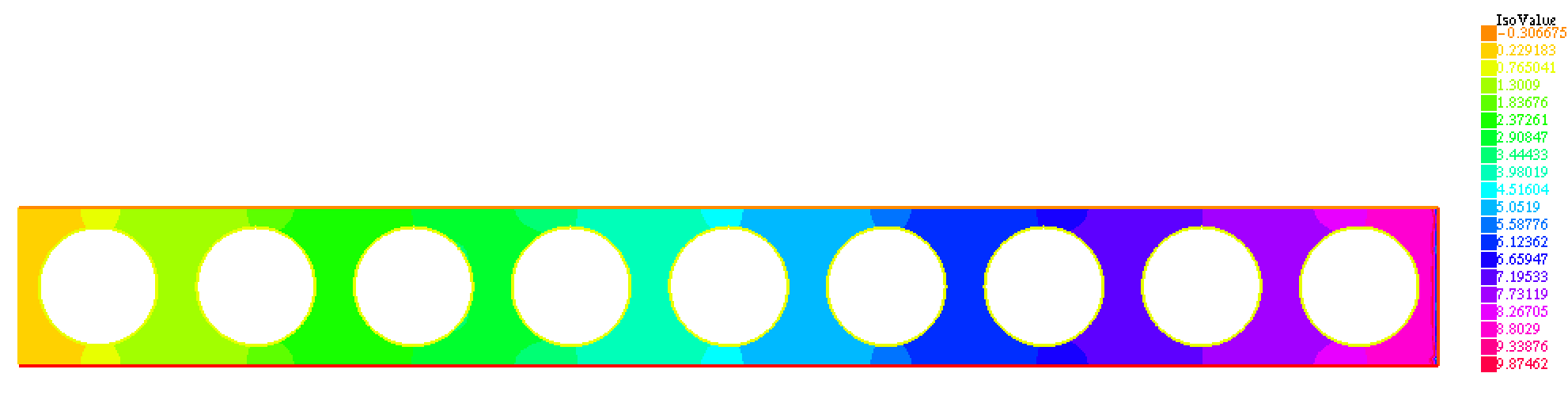}
\caption{\label{ballrow1}{\footnotesize
Navier-Stokes solution due to an horizontal pressure gradient which induces the water to flow from right to left. The color map shows the pressure. }}\end{center}
\end{figure}

\begin{figure}[htbp]
\begin{center}
\includegraphics[width=13cm]{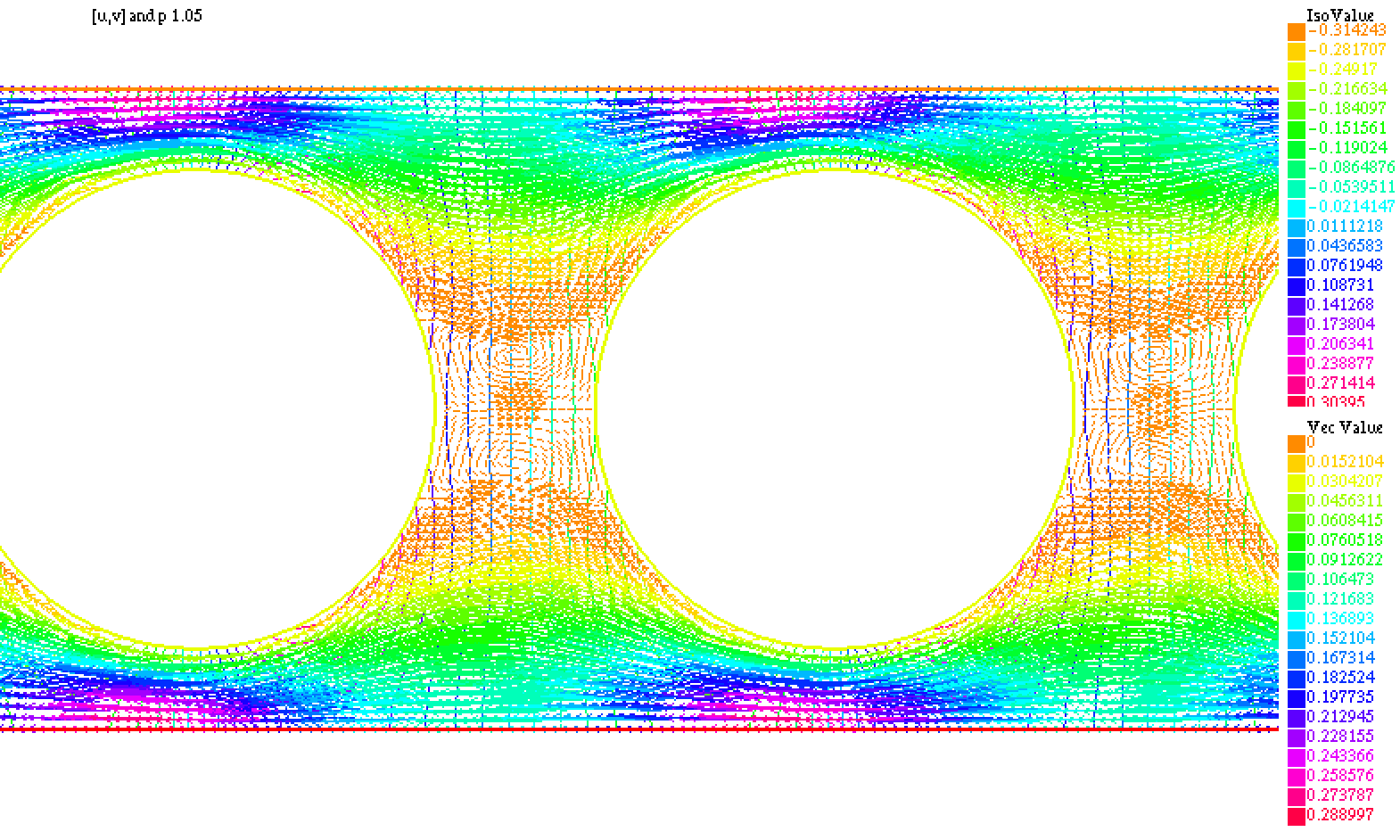}
\caption{\label{ballrow2}{\footnotesize
Navier-Stokes solution. Zoom between the second and third disk from the left showing the velocity vectors and the pressure lines..}}
\end{center}
\end{figure}

For simplicity the computation is done first in 2D, in a rectangular domain $D_p=(0,L_x)\times(0,L_y)$ with N=9 disks of radius $R=0.012$ at a distance from their neighbour $d_{ist}=0.004$ as shown on figure \ref{ballrow1}. It should be understood as being put at some point in the cylinder of figure \ref{axiveloc} perpendicular to the plane of the figure.
A stationary solution of the Navier-Stokes equations (\ref{nseq}) is computed in $D_p$ driven by a pressure gradient $\partial_x p_0=30/L_x$ and periodic boundary conditions on top-to-bottom and on left-to-right boundaries.
Results are shown on figures \ref{ballrow1} and  \ref{ballrow2} with :
\[
L_x=2 N(R+d_{ist}),~L_y=2(R+d_{ist}),~\nu=\epsilon=1/100
\]
Notice that the flow is quite periodic so that one disk only and periodic conditions is sufficient to analyse the flow.
Finally we use the velocity field computed above and solve (\ref{cconcentr}) with $\xi=0.02 $ and $\xi=0.2$.  Periodic conditions are imposed on top and bottom boundary; on the right boundary $C=1$ and on the left boundary no condition is imposed.
\begin{figure}[htbp]
\begin{center}
\includegraphics[width=12cm]{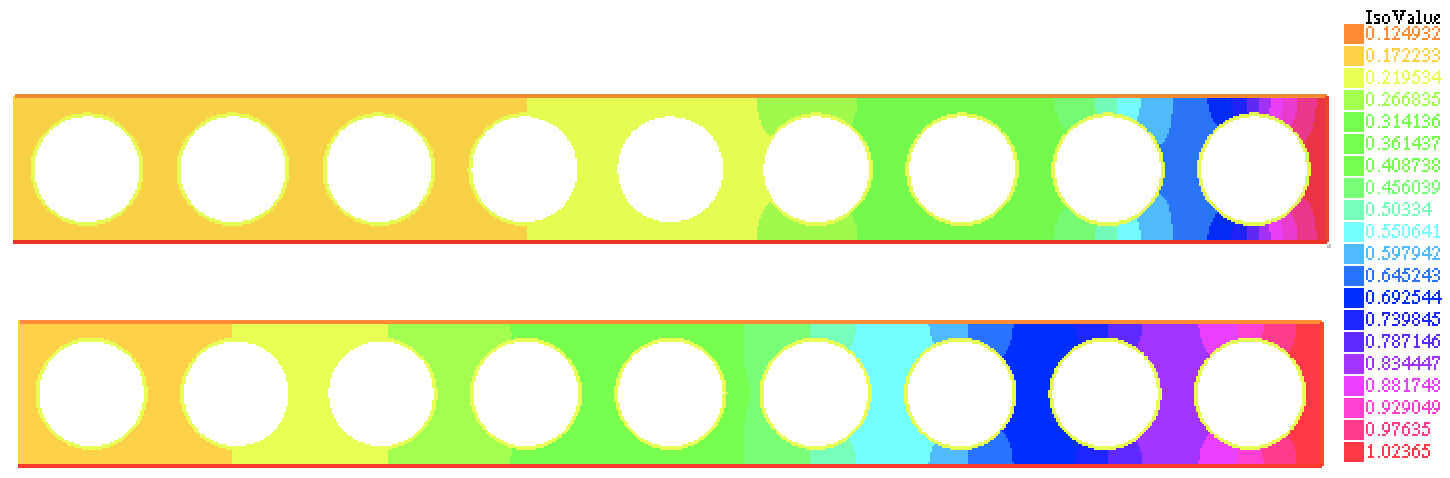}
\caption{\label{ballrow3}{\footnotesize
Color plots showing the concentration $C$ solution of (\ref{cconcentr})
when $\xi=0.2$ (top) and $\xi=0.02$ (bottom). Notice that the concentration decreases from 1 to 0.88 after the first disk when $\xi=0.02$ and from 1 to 0.65 when $\xi=0.2$.}}
\end{center}
\end{figure}
The computation shows, on figure \ref{ballrow3}  the concentration when the system has reached a stationary state, here at t=2. Notice that much of the calcite is transformed by the first balls from the right, but again much more efficiently when $\xi=0.2$.
\subsubsection{3D Computation}
\begin{figure}[htbp]
\begin{center}
\includegraphics[width=4cm]{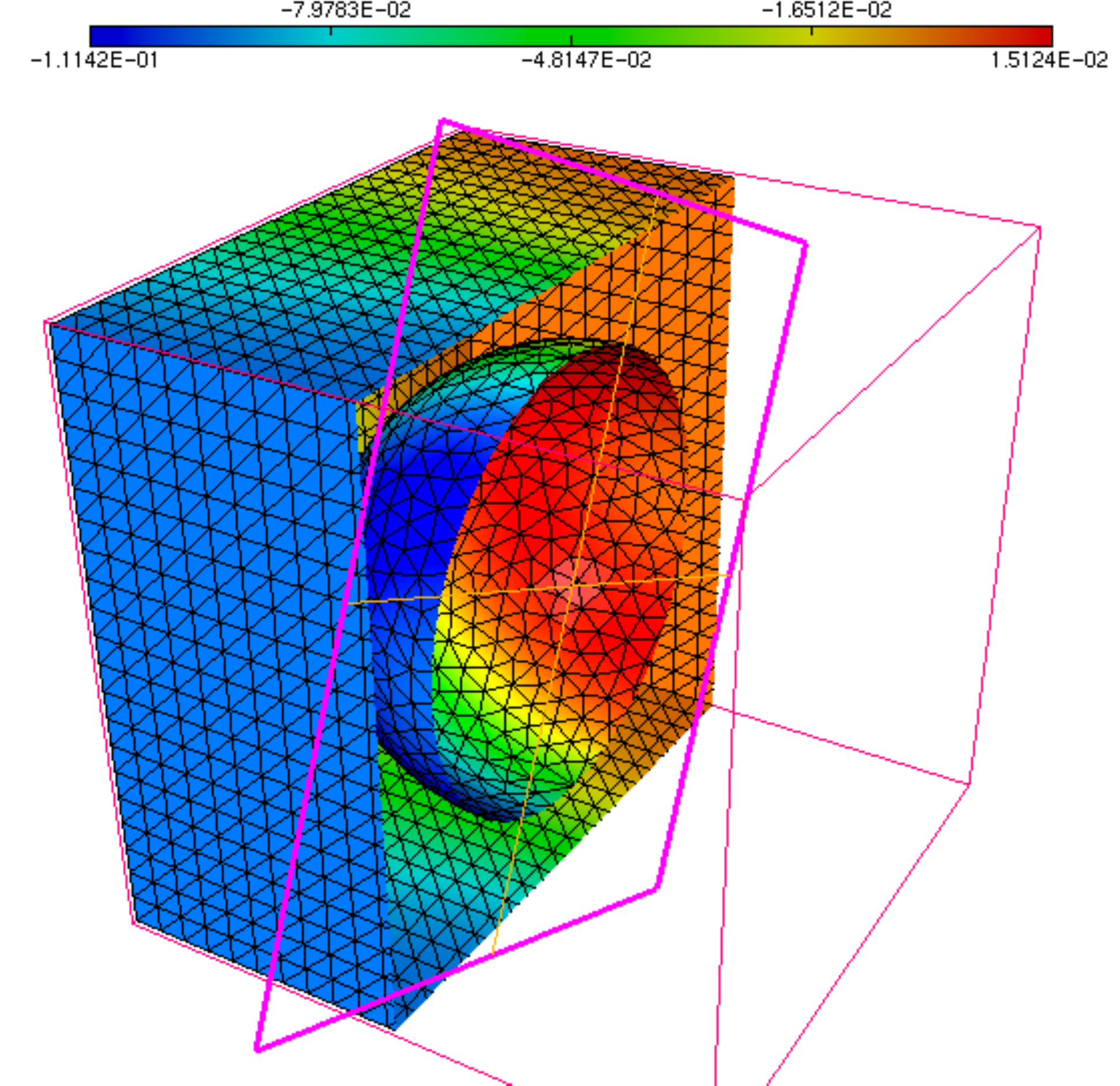}
\includegraphics[width=4cm]{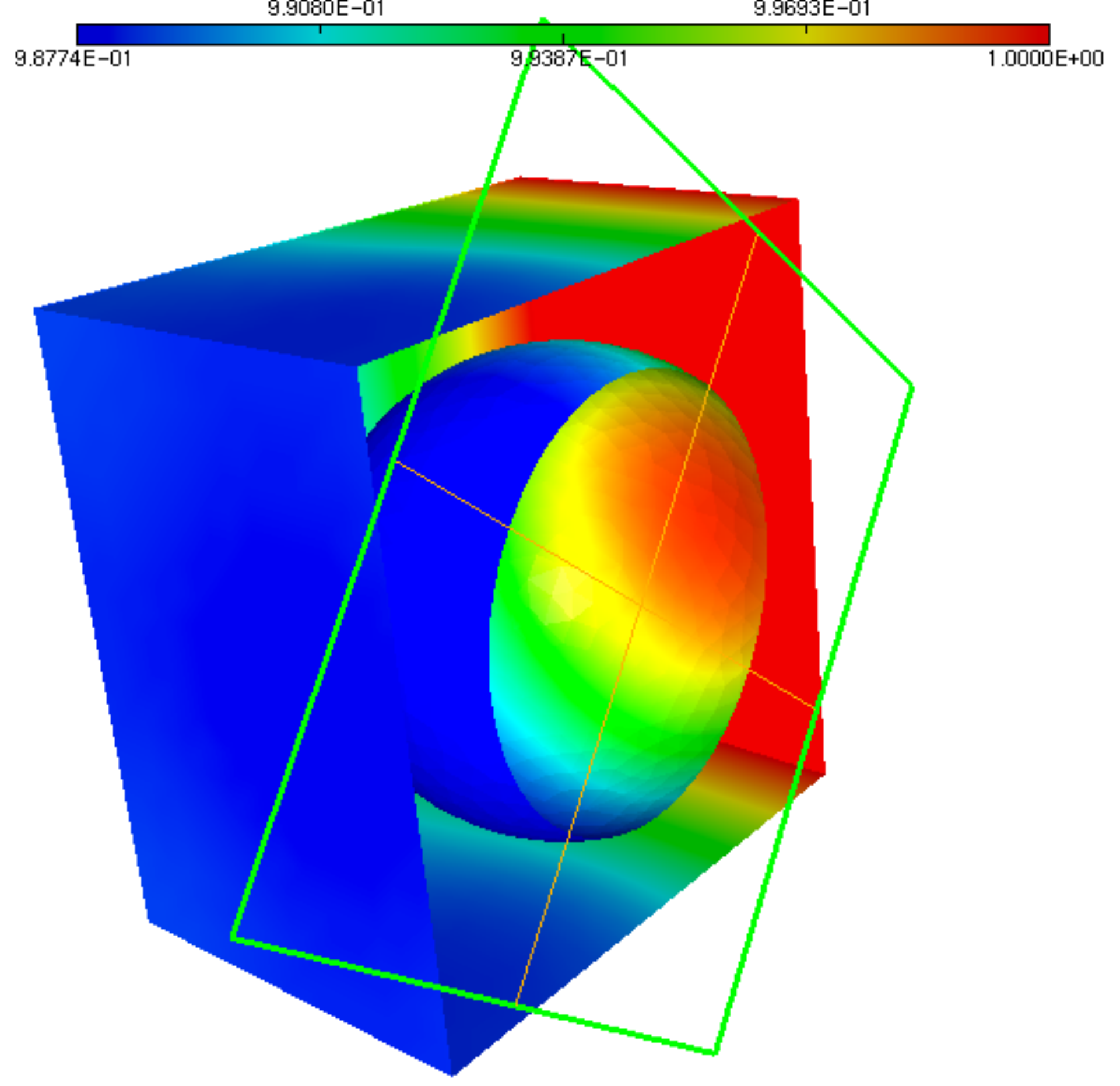}
\includegraphics[width=3.5cm]{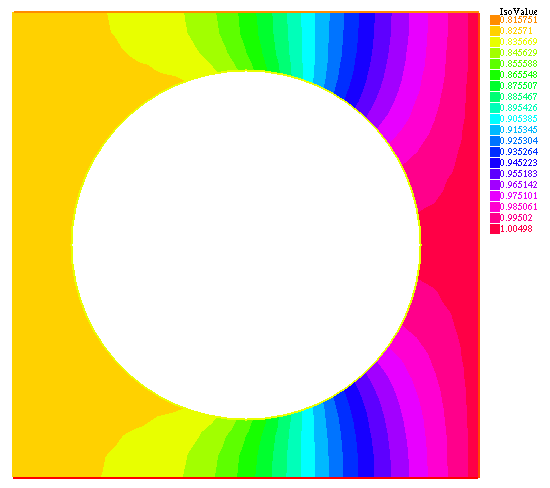}
\caption{\label{troisd} {\footnotesize
3D simulation. Left: Pressure in the fluid for a Navier-Stokes flow driven by a difference of pressure from right to left. Right: concentration of calcite, solution of (\ref{cconcentr}). Here $\xi=0.2$. Notice that the concentration decreases from 1 at the inlet to 0.987 at the outlet boundary, while for the same geometry and pressure difference but in 2D disk the calcite concentration decreased from 1 to 0.81 (lower plot).
}}
\end{center}
\end{figure}

Ideally we need to solve the same problem in 3D, namely an array of balls in a parallelogram with periodic conditions on the sides and a flow driven by a pressure gradient.  However the 2D simulations above show that we can gather already a fair amount of information by solving the flow and the equation for the calcite around one ball only provided periodic conditions are applied.
\\
The following is a computation of the Navier-Stokes equations coupled with (\ref{cconcentr}) with one ball only of radius $R=0.012$ in a cube of size 0.026 with periodic conditions on the velocities and driven by a pressure gradient equal to -3/0.026, a value similar to the 2D case. The concentration is imposed equal to 1 on the side of the cuber where the flow enters and no condition on other boundaries, except (\ref{cbdy}) on the ball. Other parameters are as in the 2D computation above, $\nu=\epsilon=0.01, ~\xi=0.2$.
\\
Results are shown on figure \ref{troisd}. The plot of the iso-surface of the concentration seems to indicate that a 3D ball is  less effective than a cylinder (2D disk) to eliminate the calcite; the calcite concentration is shown  at t=1. It also shows that 2D computations do not reflect reality and a massive 3D computation needs to be done for reliable numbers.
\\\\
To conclude, all computations have demonstrated a positive influence of the ceramic balls on the formation of scale in the sense that strongly polarised nuclei give rise to less surface tension energy which in turn implies a larger $\xi$ and hence less concentration of calcite in the water passing near the ceramic balls.

\section*{Conclusion}

The effect of an array of ceramic balls on hard water has been studied. A number of facts have been found which could possibly explain the beneficial effect of the balls on the growth of calcite crystals of \texttt{CaCO}$_3$.\

A model to compute the electric field at the surface of the balls has been given. It has been found that the number of polarized nuclei which contribute to scale formation is considerably smaller than in natural water.

 Then a two dimensional macroscopic model has been investigated numerically showing that the effect of the ceramic balls could be reproduced in configurations studied experimentally and industrially.
\\
The general conclusion is  that the ceramic balls induce a polarization in the calcite particles which decrease the surface tension energy of the crystal  in the vicinity of the ceramic balls.
\paragraph{Acknowledgements}
The authors are grateful to S.Ikawa for bringing them first about what and where the essential problem in this paper is. They would like to thank  H.Fujita for the useful discussions related with this paper.    They also would like to thank  R. Inaba ( MSSA S.A.S.)for furnishing data on thermodynamic properties of calcium carbonate and the surface physics and chemistry.  Finally the first author would like to thank to  H. Suito for his long standing scientific support.
\\
The first author's contribution to this paper is for the modeling while the second author has concentrated on the numerical simulations.

\section{Appendix: Ginzburg-Landau model for the phase change}

Crystallization experiments were performed with the standard (NH$_4$)$_2$CO$_3$  diffusion method \cite{kawarada}.

The aspects of the sediments  in the absence of ceramic balls and in the presence of them were observed with an electron microscope \cite{kawarada}. The former case shows that the sediments are parallelipedic of size $10^{-4}$(m) in average.  In the latter case  the sediments the size o is about $5\cdot  10^{-6}$(m) in average. 

The electric field caused near the surface of the ceramic balls seems to stop the growth of the sediments at roughly $1/20$ the length of the rectangular shape. 
In this section we try to clarify the mechanism for this difference by using a Landau-Ginzburg model \cite{doi} with an electric field. 

Here we discuss the effects of  exposing to ceramic balls the crystal particle of calcite, in the later period after crystallization.   The important parameter is the density of material $\psi=\psi \left( r, t \right)$ .  In the Ginzburg-Landau model, free energy per unit volume is approximated by a polynomial of degree 4 (see \cite{doi})
\begin{eqnarray}
H=\frac{1}{2}\tau \psi^2 + \frac{1}{4}u_0 \psi^4 - h\psi - \frac{1}{3} v_0 \psi^3+ \frac C2\left| \mbox{grad} \psi \right| ^2 \; (\mbox{J/m}^3)
\end{eqnarray}
where $\tau$ is a negative constant and $u_0$ and $v_0$ are positive constants;
as before $h$ is electric polarization energy for a single molecule. 
The last term in $H$ models the free energy density in case of non uniform density. $C$ is some positive number.

These constants imply that $H$ is a double well  potential with a meta-stable state and a stable state. These states correspond to the co-exitence of two phases, a liquid and a solid phase.

We suppose that the crystal particles are spheres. Then spherical symmetry is assumed and the domain in which $H$ is minimized is $\Omega=\{x\in R^d~:~|x|<R^*\}$.  Here
$R^{*}$ is a sufficiently large and positive number and $d=1,2$ or $3$ is the dimension of the model .  

An equation for $\psi$ is derived by saying that its evolution is given by  the derivative of $H$ in $\psi$, 
\begin{eqnarray}\label{psigrow}
\frac{d\psi}{dt}&=&-L \partial_\psi H 
=L\left(C_h \Delta\psi - \tau\psi+v_0 \psi^2 - u_0 \psi^3 \right) + L h. \label{eqn:eqm}
\end{eqnarray}
$L$ is a diffusion coefficient, and $Lh$ is an external force added to the system according to the fluctuation dissipation theorem {\cite{doi}}.

It is known that a solution of the equation exists in an appropriate Sobolev space, global in time \cite{lions}; it implies also smoothness of the solution under suitable initial and boundary conditions.

We assume that the initial condition for $\psi$, depends only on the radial coordinate $r\in(0,R^*)$:
\begin{eqnarray}
	\psi (r,0) = -K \tanh \left(r-R_0 \right) \sqrt{\frac{2}{\tau C_h}}, \;\; 0 < r < R^{*}. \label{eqn:psiIC}
\end{eqnarray}  
This choice (\ref{eqn:psiIC}) as the initial condition is appropriate to show that the root of the solution $\psi = 0$ defines the boundary between two phases. $K$ is some positive constant.
$R_0$ is a sufficiently small number. 
At $R^*$ we impose, 
\begin{eqnarray}
\psi (R^*,t) = -K.
\end{eqnarray}
Note that in 1D, when $v_0 = h = 0$, $R^*=+\infty$, (\ref{psigrow}) becomes, in the asymptotic in time regime,
\begin{eqnarray}
 \tau \psi +u_0 \psi^3 -C_h \psi'' = 0.
\end{eqnarray}
and the solution, denoted  $\psi_{int}$, is
\begin{equation}\label{tan}
	\psi_{int}= -K  {\tanh}\left(x\sqrt{\frac{2}{\tau C_h}}\right) .
\end{equation}
\begin{figure}[htbp]
\centering
\includegraphics[width=5cm]{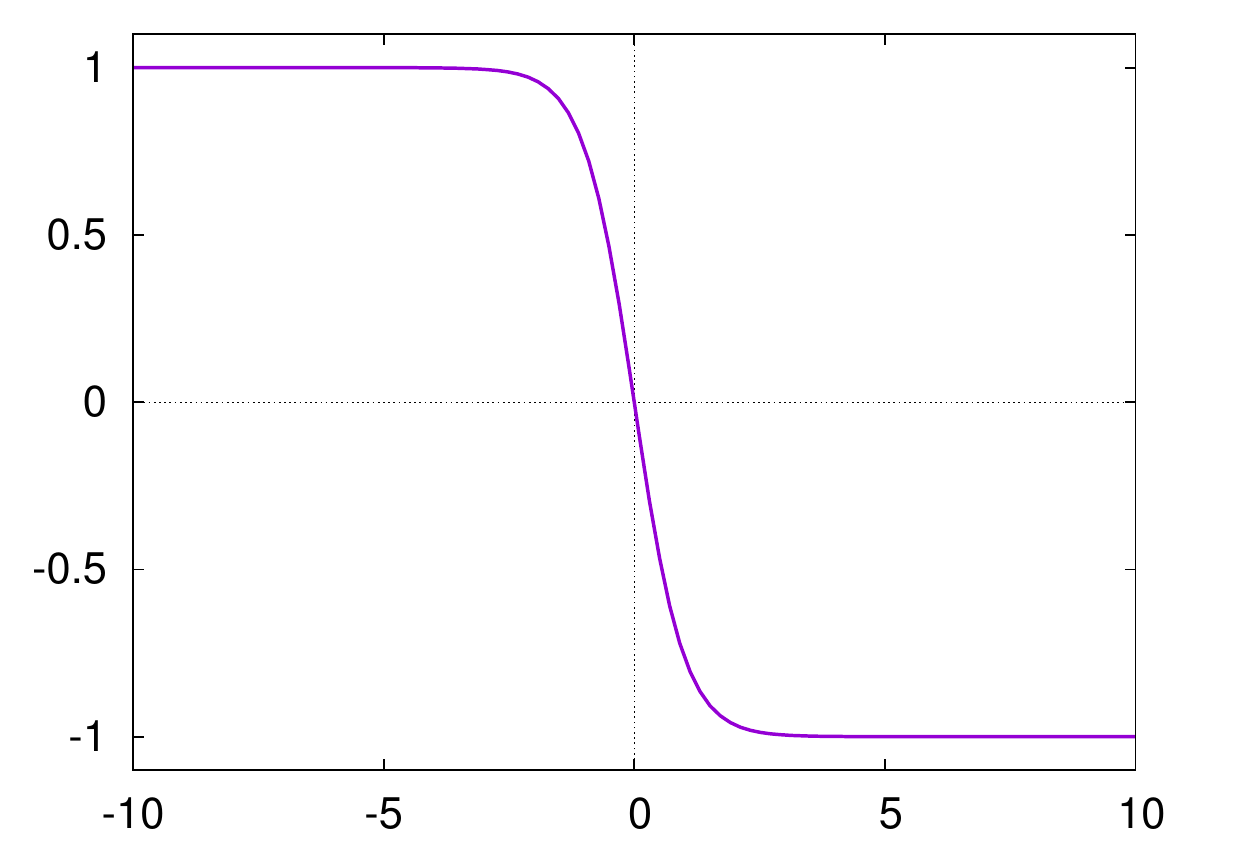}
\caption{\label{psiint} {\footnotesize Plot of $x\mapsto \psi_{int}(x)$ of (\ref{tan}) showing its fast variation near the root R}}
\end{figure}
The variation of $\psi_{int}$ near  the root, denoted by $R$, of the equation $\psi = 0$ is large compared with the variation elsewhere, as shown in figure \ref{psiint}.
\\\\
The surface tension energy $\gamma$, defined in \cite{doi}, is
\begin{eqnarray}
\gamma = C_h\int_0^\infty \left( \frac{d\psi_{int}}{dx} \right)^2 dx
\end{eqnarray}
In the general case with spherical symmetry (\ref{eqn:eqm}) reduces to 
\begin{eqnarray}
\frac{d\psi}{dt}=L\left\{C_h\left(\psi''+\frac{2}{r} \psi' \right) -\tau \psi+v_0 \psi^2  - u_0 \psi^3 +h \right\}
\end{eqnarray}
where the ${}'$ indicates a derivative with respective to $r$.

Define $V_n= dR/dt$ and replace $\psi$ by $\psi_{int}$.
 Let $x$ be the distance to the R.
Then up to first order terms in $\delta t$, it holds 
\begin{eqnarray}
\lefteqn{\frac{\psi_{int}(x) - \psi_{int}\left(x-V_n \cdot \delta t \right)}{\delta t}
= V_n \cdot \psi_{int}' } \nonumber \\
 &=& L\left \{  C_h\left( \psi_{int}'' + \cdot \psi_{int}' \cdot \frac{2}{R} \right) + h+ v_0 \psi_{int}^2 - \tau \psi_{int} - u_0 \psi_{int}^3 \right\}
\end{eqnarray}

Multiplying  both sides by $\psi_{int}'$ and then integrating in $x$, there holds
\begin{eqnarray}
\gamma \cdot V_n \cdot \frac{1}{C_h} = L\left( \gamma \cdot \frac{2}{R} - 2K \cdot h -v_0 \frac{2}{3} K^3 \right)
\end{eqnarray}
By (\ref{tan})
\[
\int \psi_{int}' \cdot dx = -2K \hbox{ and } \int \psi_{int}^2 \psi_{int}' dx = -\frac{2}{3} K^3.
\]
An equation for the evolution of $R$ follows:
\begin{eqnarray}
\frac{dR}{dt}&=& 2LC\left( \frac{1}{R} - \frac{1}{R_c} -\frac{1}{R'_c}\right), \;\;R(0)=R_0 \nonumber \\
\label{eqn:egR}
\end{eqnarray}
where  $\displaystyle R_c = \frac{\gamma}{ h K}$ is interpreted as the radius of sediments in the presence of ceramic spheres and $\displaystyle R'_c= \frac{3\gamma}{ v_0 K^3}$ is the radius of sediments in the absence of ceramic spheres.

Thus the time-asymptotic value of $R$ is $R_c/(1+R_c/R'_c)$ which is close to $R_c$ because $R_c/R'_c$ is small.  Our measurements \cite{kawarada} confirm that the radii of the crystals are comparable to $R_c=1.54~10^{-5}$.
The above mentioned discussion seems to support the validity of the contact model.

\section{Numerical solution of the Electro-Chemical Equation (\ref{system}) }

\begin{figure}[htbp]
\begin{center}
$$\includegraphics[width=6cm]{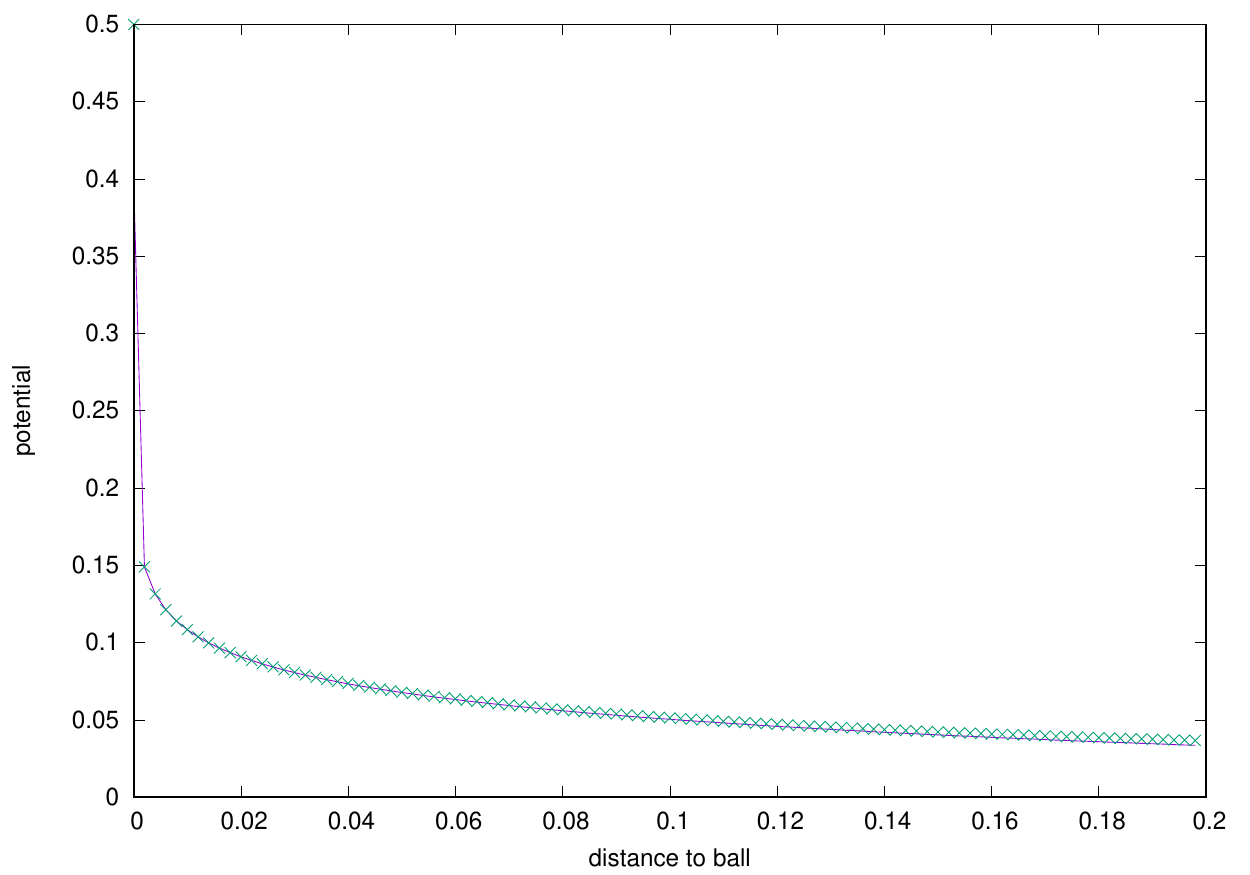} ~~\includegraphics[width=6cm]{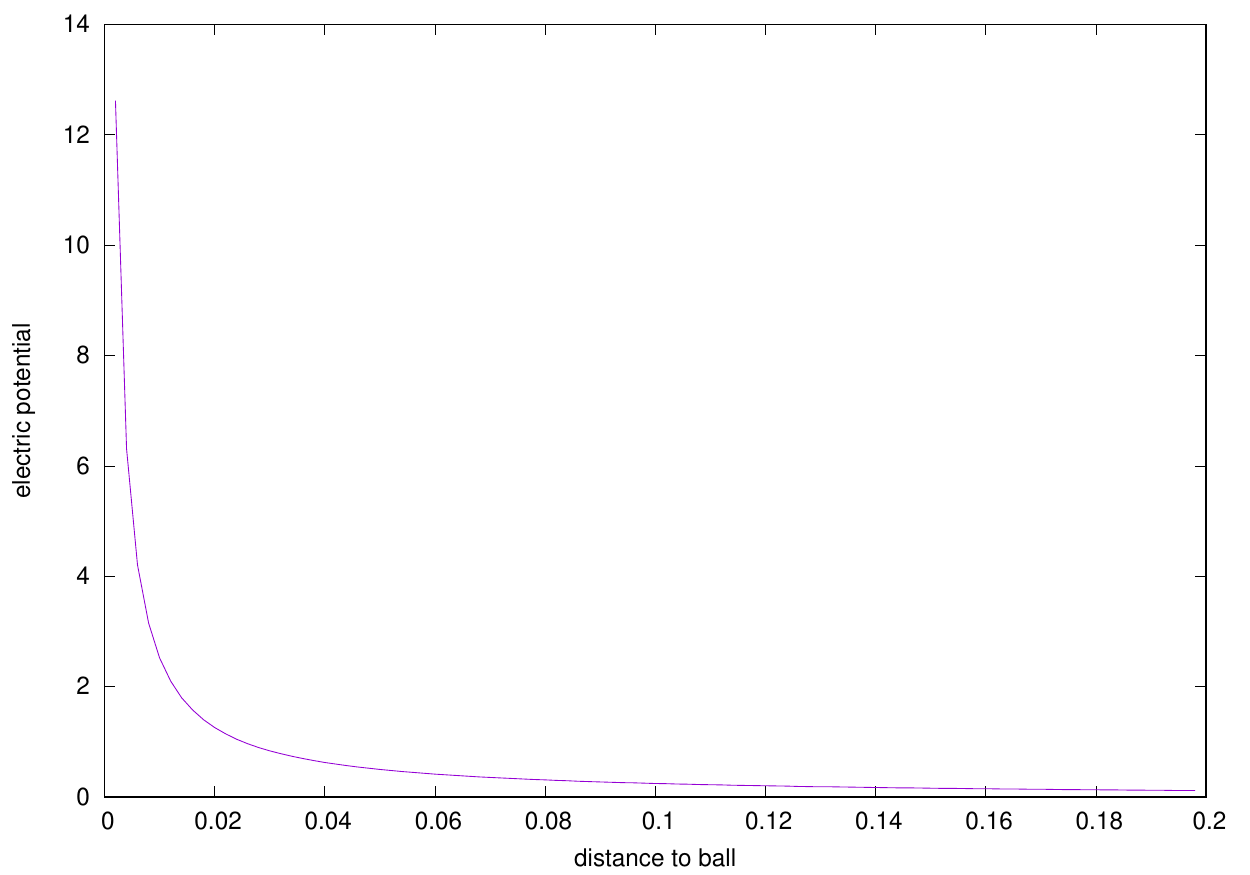}$$
\caption{{\footnotesize Left: Plot of $-\phi$ versus the radial distance to the ceramic ball. The solid line is the result of   (\ref{cosh}) solved by a numerical method. The crosses are  obtained by (\ref{phimaple}). Right: Plot of the electric field $E$ versus the radial distance; the first few points near the ball (x=0) are not plotted because the values are too big. }}
\label{psi1}
\end{center}
\end{figure}

Returning to (\ref{getE}),  multiplying by $\sinh\psi$, we obtain
\begin{equation}\label{cosh}
u:=\cosh(\psi),~~u'=-(u-1)\sqrt{2\gamma(u+1)} ,~~ u(R_b)=\cosh(b\phi_1).
\end{equation}
This ODE has a closed form solution in the complex plane (bold face letters indicates  complex values and ${\bf i}=\sqrt{-1}$):
\begin{eqnarray}
 x&=&-\frac1{\sqrt\gamma}\tanh^{-1}\sqrt\frac{{\bf u}+1}2-C{\bf i} \hbox{ with $C$ such that }u(R_b)=b\phi_1, \hbox{i.e.}
 \cr
~~ {\bf u}-1&=&2\left(\tanh[\sqrt\gamma({\bf x}+C{\bf i})]\right)^2
\cr&&
 =2 \frac{(1+T^2)^2\tanh^2(x\sqrt\gamma)-T^2(1+\tanh(x\sqrt\gamma))^2}{(1+T^2\tanh^2(x\sqrt\gamma))^2} +{\bf i} (..)
\end{eqnarray}
where $T=\tan(C\sqrt\gamma)$ is extremely large because $\cosh(b\phi_1)\sim O(10^{13})$. Hence
\begin{equation}\label{phimaple}
u(x)\approx 1+ \frac2{\tanh^2(x\sqrt\gamma)},~~ \phi(x)\approx \frac1b\cosh^{-1}(1+ \frac2{\tanh^2(x\sqrt\gamma)}).
\end{equation}
Figure \ref{psi1} shows on the left $x\mapsto\phi(R_b+x)$. The solid line is the result of   (\ref{cosh}) solved by a numerical method when $C_0=0.017$ mol m$^{-3}$ and \texttt{[pH]}=7. The crosses are  obtained by (\ref{phimaple}).  The  electric potential $E$ with the distance to the ball is reported on the left of Figure  \ref{psi1}; however the points nearer to the balls are not plotted because at the ball the computed field is  $0.7902\cdot 10^7$.  It is smaller than the value calculated in section \ref{sec3} by several order of magnitude because very near the ball the numerical grid is not fine enough. 
Finally we studied numerically the dependency of $E$ on the dimension parameter $d$ by solving numerically (\ref{system})  and found no difference in any of the previous computations, whether the modelling be for a sphere, a cylinder or a plane.

\newpage


%




\end{document}